\pdfoutput=1
\documentclass[preprint,prl]{revtex4}

\usepackage{graphicx}% Include figure files
\usepackage{dcolumn}% Align table columns on decimal point
\usepackage{bm}% bold math

\begin{document}
%\preprint{APS/123-QED}

%\draft
% ****** Start of file apssamp.tex ******
%
%   This file is part of the APS files in the REVTeX 4 distribution.
%   Version 4.0 of REVTeX, August 2001
%

\title{Quantum Fluctuations in a Cavity QED System with Quantized Center Of Mass Motion}

%>>>> The author is responsible for formatting the
%  author list and their institutions.  Use  \skiplinehalf
%  to separate author list from addresses and between each address.
%  The correspondence between each author and his/her address
%  can be indicated with a superscript in italics,
%  which is easily obtained with \supit{}.

\author{J. R. Leach$^{1,3}$, M. Mumba$^2$, and P. R. Rice$^3$}%
 \email{ricepr@muohio.edu}
\affiliation{ $^1$Department of Radiology and Biomedical Imaging, University of California San Francisco, San Francisco, CA 94143  \\
$^2$Department of Physics, University of Arkansas,
Fayetteville, AR 72701\\
$^3$ Department of Physics, Miami University, Oxford,
Ohio 45056 {\email{ricepr@muohio.edu}}%
}

\date{\today}% It is always \today, today,

%%%%%%%%%%%%%%%%%%%%%%%%%%%%%%%%%%%%%%%%%%%%%%%%%%%%%%%%%%%%%
\begin{abstract}
We investigate the quantum fluctuations of a single atom in a
weakly driven cavity, where the center of mass motion of the atom
is quantized in one dimension. We present analytic results for the
second order intensity correlation function $g^{(2)}(\tau)$ and
the intensity-field correlation function $h_{\theta}(\tau)$, for
both transmitted and fluorescent light for weak driving fields. We
find that the coupling of the center of mass motion to the
intracavity field mode can be deleterious to nonclassical effects
in photon statistics; less so for the intensity-field correlations.

\end{abstract}

  \maketitle
%%%%%%%%%%%%%%%%%%%%%%%%%%%%%%%%%%%%%%%%%%%%%%%%%%%%%%%%%%%%%
\section{Introduction}
\label{sect:intro}  % \label{} allows reference to this section
Since the mid 1970's, quantum opticians have been investigating
explicitly nonclassical states of the electromagnetic field, and
ways to determine if a field state {\it is} nonclassical. These
types of states are ones for which there is no underlying
non-singular probability distribution of amplitude and phase, or
more technically, they exhibit a positive definite
Glauber-Sudarshan P distribution. Much work has focused on photon
antibunching, sub-Poissonian photon statistics, quadrature
squeezing, and entangled atom-field states\cite{nonclass}. The
generation of such light fields may have applications in quantum
information processing, atomic clocks, and fundamental tests of
quantum mechanics, for example. One system that has long been a
paradigm of the quantum optics community is a single-atom coupled
to a single mode of the electromagnetic field, the Jaynes-Cummings
model\cite{JC}. In practice the creation of a preferred
field mode is accomplished by the use of an optical resonator.
This resonator generally has losses associated with it, and the
atom is coupled to vacuum modes out the side of the cavity leading
to spontaneous emission. Energy is put into the system by a
driving field incident on one of the end mirrors. The
investigation of such a system defines the subfield of cavity
quantum electrodynamics\cite{CQED}.  The
presence of the cavity can also be used to enhance or reduce the
atomic spontaneous emission rate\cite{CQED}. This system has also
been studied extensively in the laboratory, but several practical
problems arise.\cite{expt1,expt2,thy} There are typically many atoms in the cavity at
any instant in time, but methods have been developed to load a
cavity with a single atom. A major problem in experimental cavity
QED stems from the fact that the atom(s)are not stationary as is
often assumed by theorists. The atoms have typically been in an
atomic beam originating from an oven, or perhaps released from a
magneto-optical trap. This results in inhomogeneous broadening of
the atomic resonance from Doppler and/or transit-time broadening.
Using slow atoms can reduce these effects, but the coupling of the
atom to the light field in the cavity is spatially dependent, and
as the atoms are in motion, the coupling is then time dependent;
also different atoms see different coupling strengths.

With greater control in recent years of the center of mass motion
of atoms, developed by the cooling and trapping community,
preliminary attempts have been made to investigate atoms trapped
inside the optical cavity\cite{Cool}. The recent demonstration of
a single atom laser is indicative of the state of the art
\cite{HJKSAL}. In this paper we consider a
single atom cavity QED system with the addition of an external
potential, provided perhaps by an optical lattice, and study the
photon statistics and conditioned field measurements of both the
transmitted and fluorescent fields. We seek to understand (with a
simple model at first) how the coupling of the atom's center of
mass motion to the light field affects the nonclassical effects
predicted and observed for a stationary atom.

The system we consider is shown schematically in Fig. 1.
\begin{figure}
   \begin{center}
   \begin{tabular}{c}
   \includegraphics[height=4cm]{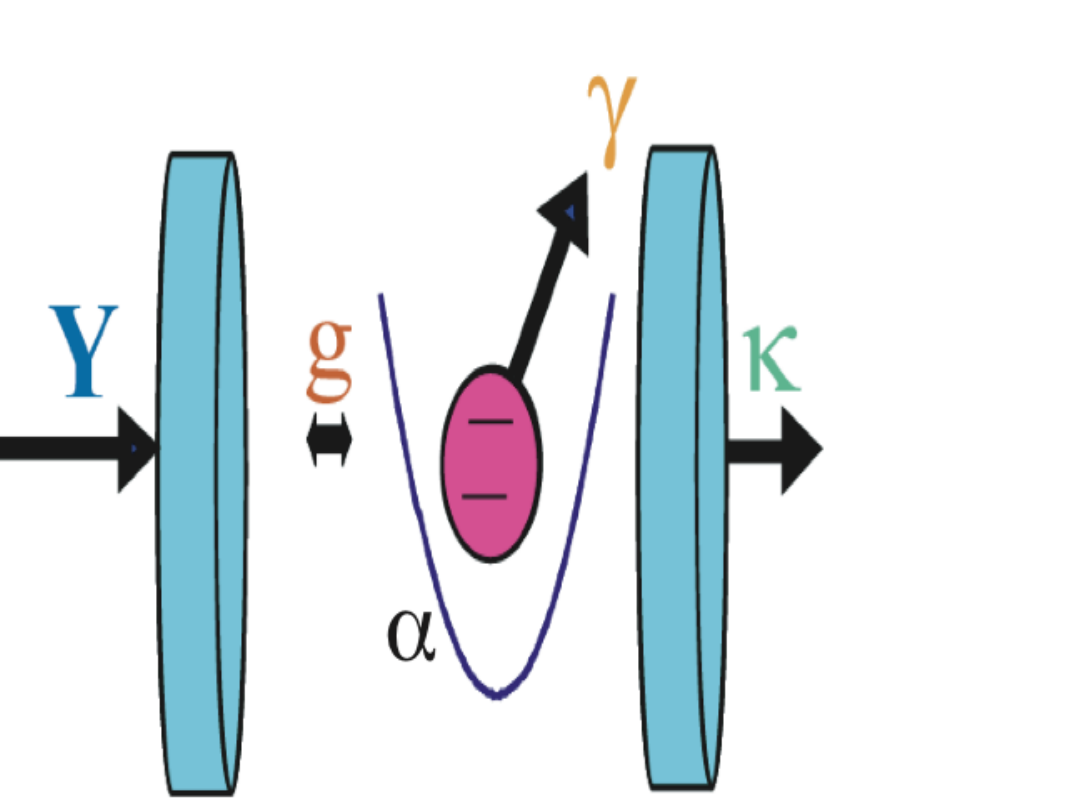}
   \end{tabular}
   \end{center}
   \caption[example]
%>>>> use \label inside caption to get Fig. number with \ref{}
%   { \label{fig:example}
{Single atom in a weakly driven optical cavity with an external
potential}
   \end{figure}We
utilize the quantum trajectory method in which the system is
characterized by a wave function and non-Hermitian Hamiltonian
\begin{eqnarray}
|\psi _c(t)\rangle &=&\sum\limits_{n,l=0}^\infty \left(
C_{n,l,g}(t)e^{-iE_{n,l,g}t}|n,l,g\rangle \right. \nonumber\\&&
\left. +C_{n,l,e}(t)e^{-iE_{n,l,e}t}|n,l,e\rangle \right) \label{psi}\\
H&=& {{p^2}\over{2m}}+V_{ext}-i\kappa a^\dagger a
-\imath{{\gamma}\over2} \sigma_+\sigma_- \nonumber
\\&&+ \imath \hbar Y(a^{\dagger }-a)+ \hbar g(\vec{r})\;(a^\dagger
\sigma _-+a\sigma _+)
\end{eqnarray}
where we also have collapse operators
\begin{eqnarray}
{\cal C}&=&\sqrt{\kappa} a\\ {\cal
A}&=&\sqrt{{\gamma}\over2}\sigma_-.
\end{eqnarray}
associated with photons exiting the output mirror and spontaneous
emission out the side of the cavity. The indices $e(g)$ indicate
the atom in the excited (ground) state, $n$ is the photon number,
and $l$ is a quantum number associated with the presence of bound
states of the external potential. We have the usual creation ($a$)
and annihilation ($a^{\dagger}$) operators for the field, and
Pauli raising and lowering operators $\sigma_{\pm}$ for the atom.
The bare energies are $E_{n,l,g}=\hbar(n\omega +\Omega_l)$ and
$E_{n,l,e}=\hbar((n+1)\omega +\Omega_l)$, where the $\Omega_l$ are
the discrete, bound, states of the external potential. The
classical driving field (in units of photon flux) is given by $Y$.
We take the external potential in which the atom is trapped to be
harmonic along the cavity axis, $V_{ext}=\alpha (z-z_0)^2/2$,
which could be appropriate for a 1-D optical lattice inside the
cavity. We ignore the generally weak transverse dependence of the
atom-field coupling, $g(\vec{r})\rightarrow g(z)=g_m f\left(
k(z-z_0)\right)$, with the maximum coupling given by
$g_m=\mu_{eg}\sqrt{\omega/2\hbar\epsilon_0V}$and
$f\left(k(z-z_0)\right)$is the cavity field mode function. Here $\mu_{eg}$ is the dipole transition matrix element, and $V$ is the 
volume of the cavity mode. We
assume for simplicity that the bottom of one of the lattice wells
coincides with an antinode of the cavity field. Following the treatment
of Kimble and Vernooy,\cite{KV} and keeping only terms to $(z-z_0)^2$, we
find the non-dissipative parts of the Hamiltonian to be
\begin{eqnarray}
H&=& {{p^2}\over{2m}}+{{\alpha}\over2}(z-z_0)^2+i\hbar
E(a^{\dagger }-a)\nonumber \\&&+\hbar
g_m\left(1+{{(z-z_0)^2}\over{2\eta^2}}\right)\;(a^\dagger
\sigma _-+a\sigma _+)\label{Ham}
\end{eqnarray}
where the characteristic distance $\eta$ is defined by
\begin{equation}
\eta^{-1}=\sqrt{{1\over{g_m}}{{d^2g(z)}\over{dz^2}}\mid_{z=z_0}}=\sqrt{{{d^2f(z)}\over{dz^2}}\mid_{z=z_0}}
\end{equation}
For a standing wave mode, $f(k(z-z_0))=cos(k(z-z_0))$ we have
$\eta = k^{-1}=\lambda/2\pi$, where $\lambda$ is the wavelength of the cavity mode. Consider the action of this Hamiltonian on the
dressed states $\mid n, \pm \rangle=(1/\sqrt{2})\left(\mid n,g
\rangle \pm \mid n-1, e\rangle\right)$ with $n$ the number of
intracavity photons, and $e(g)$ denotes the excited (ground) state
of the atom.  As $(a^\dagger \sigma _-+a\sigma _+)\mid n, \pm,
l\rangle=\pm\sqrt{n}\mid n, \pm, l\rangle$ we have an effective
potential
\begin{equation}
V(z)={1\over2}\left(\alpha\pm{{\hbar g_m\sqrt{n}}\over{\eta^2}}\right)(z-z_0)^2={1\over2}m\Omega_{n\pm}^2\label{trappot}
\end{equation}
with an effective harmonic frequency $\Omega_{n\pm}$ defined above. In the dressed state basis, the selection rule for dipole transitions is$\Delta l=0$. It is worth noting that in a basis defined by an outer product of the atom-field dressed states and the bare vibronic levels of the external potential enumerated by the quantum number $L$, which we call the casually dressed states, the selection rule is $\Delta  L =0,\pm 2$. These arise from absorption of a photon traveling to the right (left) in the cavity, with reemission into the same direction ($\Delta L=0$), while absorption of a photon traveling in one direction and emission into the opposite direction leads to a momentum kick for the atom or $2\hbar L$, leading to $\Delta L=\pm 2 $.

 We then use the  dressed states
$\mid n, l,\pm \rangle$ where the index $l$ denotes the degree of
excitation in the vibronic states corresponding to combined
lattice/field coupling potential. Please note that these are not
the vibronic states of the optical lattice alone. An energy level
diagram is shown in Fig. 2.
\begin{figure}
   \begin{center}
   \begin{tabular}{c}
   \includegraphics[height=9cm]{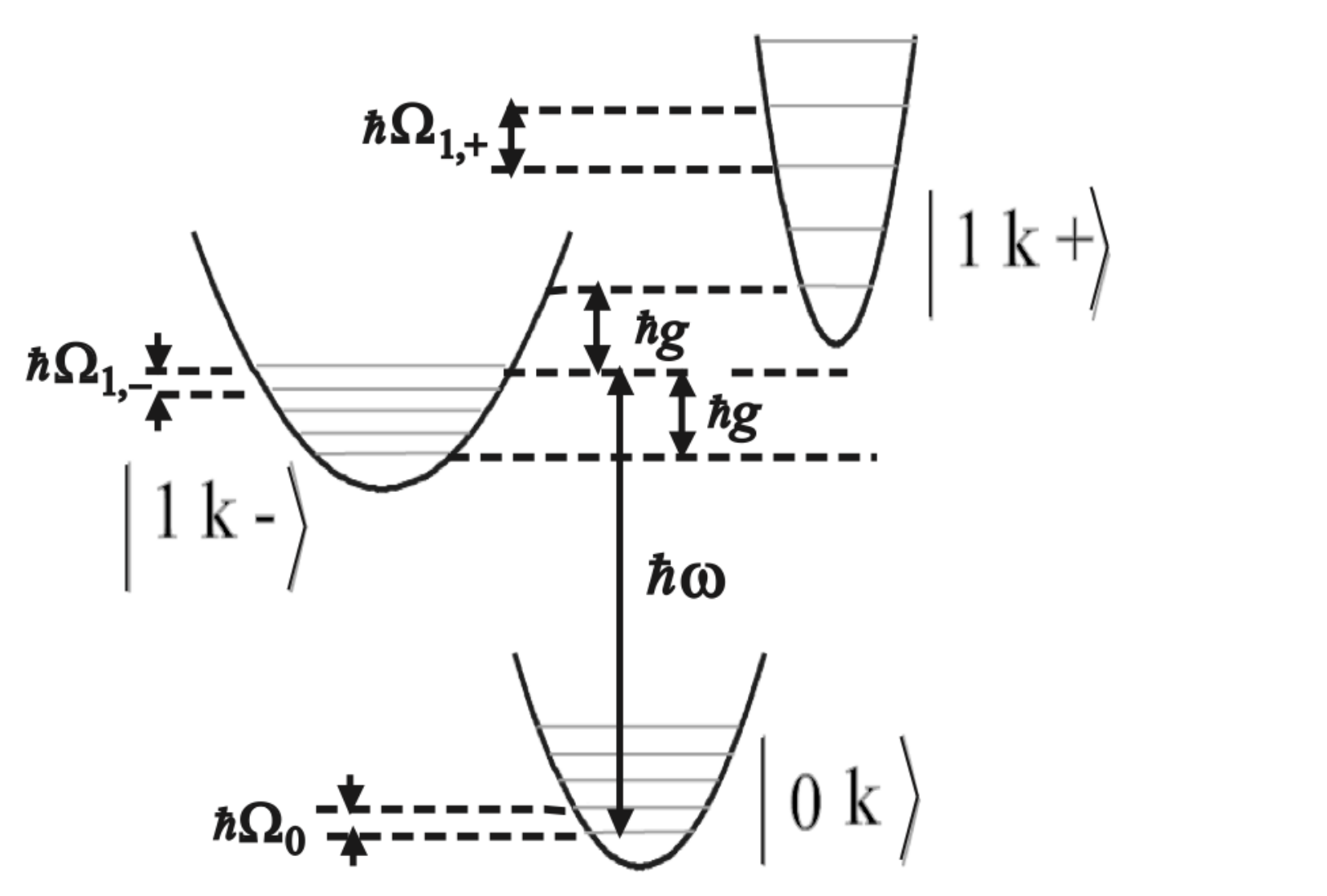}
   \end{tabular}
   \end{center}
   \caption[example]
%>>>> use \label inside caption to get Fig. number with \ref{}
%   { \label{fig:example}
{Energy level diagram}
   \end{figure}
We notice that the level spacings of the three sets of dressed
vibronic states are not equal, due to the $\pm g\sqrt{n}$ term in
the vibronic frequency.

\section{Intensity-Intensity Correlations}
 We next consider the second-order intensity
correlation function $g^{(2)}(\tau)=\langle
a^{\dagger}(0)a^{\dagger}(\tau)a(\tau)a(0)\rangle/{\langle
a^{\dagger}a\rangle}^2_{SS}$.

In the weak field limit, only states with 2 or fewer quanta of
energy are left within the basis (we must keep states with at
least \itshape two \upshape photons, as we wish examine photon coincidences).   The
limit we are considering is one in which $Y\rightarrow 0$; and we truncate our equations of motion to lowest order in $Y$. If no driving field were applied, the atom would certainly
be in the ground state, so we make the approximation that for weak
fields $C_{0,l,g} \sim 1$. With no trapping potential, one would have $C_{g,0}=1.0$; here we must specify the set of initial populations which correspond to the center of mass motion of the atom, subject to the normalization condition $\sum_l |C_{0,l,g}|^2=1.0$. The potential is taken to be of the same sign for plus and minus dressed states, which is possible by placing the lattice field at a \lq\lq magic \rq\rq frequency \cite{magic1,magic2,magic3}.The driving field is responsible for
populating the atom's excited states, and thus $C_{1, l, \pm} \sim
Y C_{0,l,g} \sim Y$. This reasoning can be continued and we determine that our scaling should be
\begin{eqnarray}\label{scale}
    C_{0,l,g} &\sim& 1 \nonumber \\
    C_{1, l, \pm} &\sim& Y \nonumber \\
    C_{2, l, \pm} &\sim& {Y}^2.
\end{eqnarray}

In the weak field limit, the one excitation amplitudes satisfy
\begin{eqnarray}
\dot{C}_{1,l,+}&=&-\left({{\gamma}\over{4}}+{{\kappa}\over2}+i\Delta_{l,1+}\right)C_{1,l,+}
-{Y\over{\sqrt{2}}}C_{0,l,g}\nonumber \\ &&-\left({{\gamma}\over{4}}-{{\kappa}\over2}\right)C_{1,l,-}\\
\dot{C}_{1,l,-}&=&-\left({{\gamma}\over{4}}+{{\kappa}\over2}+i\Delta_{l,1-}\right)C_{1,l,-}
-{Y\over{\sqrt{2}}}C_{0,l,g}\nonumber \\&&
-\left({{\gamma}\over{4}}-{{\kappa}\over2}\right)C_{1,l,+}
\end{eqnarray}

with $\Delta_{1,l\pm}=\left(\Omega_{1,\pm}-\Omega_0\right)l\pm g$,
recall the effective harmonic frequency
\begin{equation}
m\Omega_{n\pm}^2= \alpha
\pm{{\hbar g_m\sqrt{n}}\over{\eta^2}}.\end{equation} and again, we
keep only lowest order terms in the driving field $Y$. These are the
frequencies that correspond to the energy levels of the system,
the $\alpha$ term arises from the external potential, the $\pm
g\sqrt{n}$ terms from the spatial structure of the cavity mode
function and the coupling of motion in the mode to the interaction
with the driving field.

 As a first step  we note that we can solve the equations of motion
for the slowly varying population amplitudes $
\dot{D}_{n,k,\pm,}$, defined as
\begin{eqnarray}\label{Ddef}
    D_{0,l,g} &=& C_{0,l,g} \nonumber, \\
    D_{1,l,\pm} &=& C_{1,l,\pm} e^{-it(\Omega_{1, \pm}l - \Omega_0 l \pm g)}
    \nonumber,
    \\
    D_{2,1,\pm} &=& C_{2,l,\pm} e^{-it(\Omega_{2, \pm}l - \Omega_0 l \pm
    \sqrt{2}g)}.
\end{eqnarray}
We find that our
$\dot{C}_{n,l,\pm}$ equations become
\begin{eqnarray}\label{Ddot0lg}
    \dot{D}_{0,l,g} &=& \dot{C}_{0,l,g} \simeq 0 \nonumber \\,
\end{eqnarray}
\begin{eqnarray}\label{Ddot1l+}
    \dot{D}_{1,l,+} &=&-\left[\frac{\gamma}{4} +\frac{\kappa}{2} + i(\Omega_{1,+}l -\Omega_{0}l +g)\right]D_{1,l,+} \nonumber \\
                    &-& \frac{Y}{\sqrt{2}}D_{0,l,g} \nonumber \\
                    &-& \left[\frac{\gamma}{4} -\frac{\kappa}{2}\right]D_{1,l,-}
                    \nonumber,
                    \\
\end{eqnarray}
\begin{eqnarray}\label{Ddot1l-}
    \dot{D}_{1,l,-} &=&-\left[\frac{\gamma}{4} +\frac{\kappa}{2} + i(\Omega_{1,-}l -\Omega_{0}l -g)\right]D_{1,l,-} \nonumber \\
                    &+& \frac{Y}{\sqrt{2}}D_{0,l,g} \nonumber \\
                    &-& \left[\frac{\gamma}{4} -\frac{\kappa}{2}\right]D_{1,l,+}
                    \nonumber,
                    \\
\end{eqnarray}
\begin{eqnarray}\label{Ddot2l+}
    \dot{D}_{2,l,+} &=&-\left[\frac{\gamma}{4} +\frac{3\kappa}{2} + (i(\Omega_{2,+}l -\Omega_{0}l + \sqrt{2}g)\right]D_{2,l,+} \nonumber \\
                    &-& Y\left[\frac{1}{\sqrt{2}}+\frac{1}{2}\right] D_{1,l,+} \nonumber \\
                    &+& Y\left[\frac{1}{\sqrt{2}}-\frac{1}{2}\right] D_{1,l,-} \nonumber \\
                    &-& \left[\frac{\gamma}{4} -\frac{\kappa}{2}\right]D_{2,l,-}
                    \nonumber,
                    \\
\end{eqnarray}
\begin{eqnarray}\label{Ddot2l-}
    \dot{D}_{2,l,-} = &-&\left[\frac{\gamma}{4} +\frac{3\kappa}{2} + (i(\Omega_{2,-}l -\Omega_{0}l - \sqrt{2}g)\right]D_{2,l,-} \nonumber \\
                    &+& Y\left[\frac{1}{\sqrt{2}}-\frac{1}{2}\right] D_{1,l,+} \nonumber \\
                    &-& Y\left[\frac{1}{\sqrt{2}}+\frac{1}{2}\right] D_{1,l,-} \nonumber \\
                    &-& \left[\frac{\gamma}{4} -\frac{\kappa}{2}\right]D_{2,l,+}
                    \nonumber.
                    \\
\end{eqnarray}

In the weak field limit, these equations have a steady state solution
\begin{equation}
|\Psi_{ss}\rangle = \sum_{n,l}\left(D_{n,l,+}^{ss}|n,l,+\rangle +D_{n.l,-}|n,l,-\rangle\right)
\end{equation}
 The system reaches this steady state, which has a very small average photon number, $\langle a^{\dagger}a\rangle \sim Y^2 \ll 1$. In any given time step $\Delta t$ the probability of a collapse is given by $P_{cav}=2\kappa \langle a^{\dagger}a\rangle \Delta t \ll 1$. Similarly the probability of a spontaneous emission event in a time step $\Delta t$, $P_{atom}=\gamma\langle \sigma_+\sigma_-\rangle \Delta t$ is small. Eventually there is a cavity emission, or a spontaneous emission by the atom, leaving the system in the states
\begin{equation}\label{Psicond}
     |\Psi_c\rangle =\left\{ \begin{array}
                 {r@{\quad:\quad}l}
                 a|\Psi_{ss}\rangle = \frac{|\Psi_{CT}(0)\rangle}{||\Psi_{CT}(0)\rangle|^2} & Transmission \\ \sigma_-|\Psi_{ss}\rangle = \frac{|\Psi_{CF}(0)\rangle}{||\Psi_{CF}(0)\rangle|^2}
                 & Flourescence
                 \end{array} \right..
\end{equation}

In the steady state, all population amplitudes $D_{n,k, \pm}$ are
constant, and we may set all $\dot{D}_{n,k,\pm} = 0$.  Equations
\ref{Ddot1l+} and \ref{Ddot1l-} then become

\begin{equation}
  \left(%
    \begin{array}{cc}
      \frac{ \gamma}{4} +\frac{\kappa}{2} + i(\Omega_{1,+}l -\Omega_{0}l +g) & \frac{\gamma}{4} -\frac{\kappa}{2} \\
      \frac{ \gamma}{4} -\frac{\kappa}{2} & \frac{\gamma}{4} +\frac{\kappa}{2} + i(\Omega_{1,-}l -\Omega_{0}l -g)\\
    \end{array}
  \right) \left(
\begin{array}{c}
  D^{ss}_{1,l,+} \\
  D^{ss}_{1,l,-} \\
\end{array}
\right) = \left(
\begin{array}{c}
  \frac{-Y}{\sqrt{2}} \\
  \frac{Y}{\sqrt{2}} \\
\end{array}
\right)D^{ss}_{0,l,g}.
\end{equation}
 Solving for $D^{ss}_{1,l,+}$ and  $D^{ss}_{1,l,-}$, we
find
\begin{eqnarray}\label{D1lss}
D^{SS}_{1,l,+} &=& \frac{A_l(G+H_l)}{F_l H_l -G^2} \nonumber, \\
D^{SS}_{1,l,-} &=& \frac{-A_l(G+F_l)}{F_l H_l -G^2},
\end{eqnarray}
where
\begin{eqnarray}
A_l &=& \frac{Y}{\sqrt{2}}D_{0,l,g} \nonumber, \\
G &=& -\left[\frac{\gamma}{4} - \frac{\kappa}{2}\right] \nonumber, \\
H_l &=& -\left[\frac{\gamma}{4} + \frac{\kappa}{2} + i(\Omega_{1,-}l - \Omega_0l - g)\right] \nonumber, \\
F_l &=& -\left[\frac{\gamma}{4} + \frac{\kappa}{2} + i(\Omega_{1,+}l -
\Omega_0l + g)\right].
\end{eqnarray}
Using the same procedure, we may use our results for
$D^{ss}_{1,l,+}$ and  $D^{ss}_{1,l,-}$ and solve equations
(\ref{Ddot2l+}) and (\ref{Ddot2l-}) for $D^{ss}_{2,l,+}$ and
$D^{ss}_{2,l,-}$, finding
\begin{eqnarray}
 D^{SS}_{2,l,+} &=& \frac{-G\beta_{2,l} - Z_l \beta_{1,l}}{G^2 - Y_l Z_l} \nonumber, \\
 D^{SS}_{2,l,-} &=& \frac{-G\beta_{1,l} - Y_l \beta_{2,l}}{G^2 - Y_l
 Z_l}.
\end{eqnarray}
where
\begin{eqnarray}
 Y_l &=& \frac{\gamma}{4} + \frac{3 \kappa}{2} + i(\Omega_{2,+}l - \Omega_0l + \sqrt{2}g) \nonumber, \\
 Z_l &=& \frac{\gamma}{4} + \frac{3 \kappa}{2} + i(\Omega_{2,-}l - \Omega_0l - \sqrt{2}g) \nonumber, \\
 \beta_{1,l} &=& -Y\left[\frac{1}{\sqrt{2}} + \frac{1}{2}\right]D^{SS}_{1,l,+} + Y\left[\frac{1}{\sqrt{2}} - \frac{1}{2}\right]D^{SS}_{1,l,-} \nonumber, \\
 \beta_{2,l} &=& +Y\left[\frac{1}{\sqrt{2}} - \frac{1}{2}\right]D^{SS}_{1,l,+} - Y\left[\frac{1}{\sqrt{2}} +
 \frac{1}{2}\right]D^{SS}_{1,l,-}.
\end{eqnarray}

Now that the steady state values for the population amplitudes
have been calculated, our task is to solve for the time evolution
of $D_{1,l,+}(t)$ and $D_{1,l,-}(t)$. The probability of a cavity emission at time $\tau$ given that one occurred at $\tau=0.0$ is $2\kappa \langle\Psi_{CT}| a^{\dagger}a|\Psi_{CT} \rangle \Delta t$, hence we have
\begin{eqnarray}
g^{(2)}_{TT}(\tau)&=&{{\langle\Psi_{CT}| a^{\dagger}a|\Psi_{CT} \rangle}\over{\langle\Psi_{SS}| a^{\dagger}a|\Psi_{SS} \rangle}}\nonumber \\
&=&{{\sum_{n,l}n|C_{g,n,1}^{CT}(\tau)|^2}\over{\sum_{n,l}n|C_{g,n,l}^{SS}|^2}}\nonumber
\\
&=&{{\sum_{l}|C_{g,1,l}^{CT}|^2(\tau)}\over{\sum_{l}|C_{g,1,l}^{SS}|^2}}
\end{eqnarray}
where we have truncated the results to lowest order in the weak
field limit.
Similarly we have for the second order intensity correlation function for the fluorescent field is given by
\begin{eqnarray}
g^{(2)}_{FF}(\tau)&=&{{\langle\Psi_{CF}| \sigma_+\sigma_-|\Psi_{CF} \rangle}\over{\langle\Psi_{SS}| \sigma_+\sigma_-|\Psi_{SS} \rangle}}\nonumber \\
&=&{{\sum_{n,l}n|C_{e,n,1}^{CF}(\tau)|^2}\over{\sum_{n,l}n|C_{e,n,l}^{SS}|^2}}\nonumber
\\
&=&{{\sum_{l}|C_{e,0,l}^{CF}|^2(\tau)}\over{\sum_{l}|C_{e,0,l}^{SS}|^2}}
\end{eqnarray}
To facilitate solving the time evolution of the one-excitation amplitudes we write them in matrix form as
\begin{equation}\label{eqn of mot}
\dot{\vec{A}}(t) = M \vec{A}(t) + \vec{\Delta},
\end{equation}
where
\begin{eqnarray}
\vec{A}(t) &=&  \left(
\begin{array}{c}
  {D}_{1,l,+}(t) \\
  {D}_{1,l,-}(t) \\
\end{array}
\right) \nonumber, \\
M &=& \left(%
    \begin{array}{cc}
      -\left[\frac{ \gamma}{4} +\frac{\kappa}{2} + i(\Omega_{1,+}l -\Omega_{0}l +g)\right] & -\left[\frac{\gamma}{4} -\frac{\kappa}{2}\right] \\
      -\left[\frac{ \gamma}{4} -\frac{\kappa}{2}\right] & -\left[\frac{\gamma}{4} +\frac{\kappa}{2} + i(\Omega_{1,-}l -\Omega_{0}l -g)\right] \\
    \end{array}
  \right) \nonumber, \\
\vec{A}(t) &=&  \left(
\begin{array}{c}
  {D}_{1,l,+}(t) \\
  {D}_{1,l,-}(t) \\
\end{array}
\right) \nonumber, \\
\vec{\Delta} &=& \frac{Y}{\sqrt{2}} \left(
\begin{array}{c}
  -1 \\
   1 \\
\end{array}
\right) D_{0,l,g}.
\end{eqnarray}
The form of the time evolution of
$D_{1,l,+}(t)$ and $D_{1,l,-}(t)$:
\begin{equation}\label{time ev}
    \vec{A}(t) = \left(Se^{\Lambda t}S^{-1}\right)\vec{A}(0) + \left(Se^{\Lambda t}S^{-1} -
    1\right)M^{-1}\vec{\Delta}.
\end{equation}

 Without showing the details of such calculations, we
arrive at
\begin{eqnarray}\label{D1lpt}
    D_{1,l,+}(t) &=& \left[\frac{{\beta_2}'}{2 {\chi}_2}D_{1,l,+}(0) +\frac{G}{2{\chi}_2}D_{1,l,-}(0)
                          + \frac{Y D_{0,l,g}}{2 {\chi}_2 \phi \sqrt{2}}\left[-H_l{\beta}_2' + G^2 -G{\beta}_2'+GF_l\right]\right]
                          e^{\lambda_1 t} \nonumber \\
                 &+& \left[\frac{{-\beta_1}'}{2 {\chi}_2}D_{1,l,+}(0) -\frac{G}{2{\chi}_2}D_{1,l,-}(0)
                          + \frac{Y D_{0,l,g}}{2 {\chi}_2 \phi \sqrt{2}}\left[H_l {\beta}_1'- G^2 +G {\beta}_1'-GF_l\right]\right]
                          e^{\lambda_2 t} \nonumber \\
                 &+& \left[\frac{H_l + G}{\phi} \frac{Y}{\sqrt{2}} \right]
                 D_{0,l,g},
\end{eqnarray}
\begin{eqnarray}\label{D1lmt}
    D_{1,l,-}(t) &=& \left[\frac{{-\beta_1}'}{2 {\chi}_2}D_{1,l,-}(0) -\frac{\beta_1'\beta_2'}{2G{\chi}_2}D_{1,l,+}(0)
                          + \frac{Y D_{0,l,g}}{2 {\chi}_2 \phi \sqrt{2}}\left[\frac{H_l}{G}\beta_1'\beta_2' - G\beta_1' +\beta_1'\beta_2'-\beta_1'F_l\right]\right]
                          e^{\lambda_1 t} \nonumber \\
                 &+& \left[\frac{{\beta_2}'}{2 {\chi}_2}D_{1,l,-}(0) +\frac{\beta_1'\beta_2'}{2G{\chi}_2}D_{1,l,+}(0)
                          + \frac{Y D_{0,l,g}}{2 {\chi}_2 \phi \sqrt{2}}\left[-\frac{H_l}{G}\beta_1'\beta_2' + G\beta_2' -\beta_1'\beta_2'+\beta_2'F_l\right]\right]
                          e^{\lambda_2 t} \nonumber \\
                 &-& \left[\frac{F_l + G}{\phi} \frac{Y}{\sqrt{2}} \right]
                 D_{0,l,g},
\end{eqnarray}
\normalsize where
\begin{eqnarray}\label{opvars}
    \chi_1 &=& \frac{F_l + H_l}{2} \nonumber, \\
    \chi_2 &=& \frac{\sqrt{(F_l - H_l)^2 + 4G^2}}{2} \nonumber, \\
    \lambda_1 &=& \chi_1 + \chi_2 \nonumber, \\
    \lambda_2 &=& \chi_1 - \chi_2 \nonumber, \\
    \beta_1' &=& F_l -\lambda_1 \nonumber, \\
    \beta_2' &=& F_l -\lambda_2 \nonumber, \\
    \phi &=& F_lH_l -G^2.
\end{eqnarray}

We are now equipped with all the necessary information to solve
for the dynamics of our system in the weak field limit.

For the rest of the paper, we restrict ourselves to the deep
trapping limit,  where $\alpha \geq g_m \sqrt{n}/ \lambda^2$. We
may then use the binomial approximation, and define
\begin{equation}
    \Omega_{n,\pm} \approx \sqrt{\frac{\alpha }{m}}\left[ 1 \pm \frac{\hbar m g_m \sqrt{n}}{2 \eta^2 \alpha}
    \right],
\end{equation}
therefore we find $\Delta_{n,\pm} = \Omega{n,\pm} - \Omega_0$ for
$n = 1,2$ to be
\begin{eqnarray}
    \Delta_{1,+} &=& \frac{\hbar g_m \sqrt{n}}{2 \eta^2 \sqrt{m\alpha}} \nonumber,   \\
    \Delta_{1,-} &=& -\Delta_{1,+} \nonumber, \\
    \Delta_{2,+} &=& \sqrt{2} \Delta_{1,+} \nonumber ,\\
    \Delta_{2,-} &=& - \sqrt{2} \Delta_{1,+}.
\end{eqnarray}
and we can characterize everything by the one detuning
$\Delta_{1,+}$. This is analogous to the Lamb-Dicke regime in an ion trap.

 By using the well dressed states, we have a set of
equations that will have a steady-state. Recall that the quantum
number $l$ is associated with a well-dressed state, and not simply
the vibrational quantum number of the lattice potential. To solve
these equations it is necessary to specify the
amplitudes, $C_{g,0,1}(0)$ that are each of order unity. They can
be related to the initial center of mass state of the atom via $C_{g,0,l}=\langle g,0,l|\psi\rangle=\int \phi_{g,0,l}^* \Psi(x) dx$
, or simply specified.

For weak driving fields, the probability of a cavity emission in a
time $\Delta t$ is given by $P_{cav}=2\kappa \langle a^{\dagger}a
\rangle \Delta t$ is quite small, as is the probability of a
spontaneous emission, $P_{spon.em.}=\gamma \langle
\sigma_+\sigma_-\rangle \Delta t$. In this case the wave function
 attains a steady state
$|\psi\rangle_{SS}=\sum\limits_{n,l=0}^\infty
\left[{C_{1,l,+}^{SS}e^{-iE_{1,l,+}t}|1,l,+\rangle
+C_{1,l,-}^{SS}e^{-iE_{1,l,-}t}|1,l,-\rangle }\right]$.

After a photon is detected in transmission, at $t=0$ the wave
function collapses to
 \begin{eqnarray}|\psi(0)
\rangle_{Coll}&=&a|\psi \rangle_{SS}/|a|\psi\rangle_{SS}|\\
&=&\sum\limits_{n,l=0}^\infty
\left(C_{g,n,l}^{Coll}(t)e^{-iE_{g,n,l}t}|g,n,l\rangle
\right.\nonumber
\\&&\left. +C_{e,n,l}^{Coll}(t)e^{-iE_{e,n,l}t}|e,n,l\rangle
\right)
 \end{eqnarray}.
The initial value of the one-photon amplitudes of the collapsed
state are related to the steady state two-photon amplitudes
 \begin{eqnarray}
C_{g,1,l}^{Coll}(0)&=&{{\sqrt{2}C_{g,2,l}^{SS}}\over{\sum_{n,l}\left[2{\mid
 C_{g,2,l}^{SS}\mid}^2+{\mid C_{e,1,l}^{SS}\mid}^2\right]}}\\
C_{e,0,l}^{Coll}(0)&=&{{C_{e,1,l}^{SS}}\over{\sum_{n,l}\left[
2{\mid
 C_{g,2,l}^{SS}\mid}^2+{\mid C_{e,1,l}^{SS}\mid}^2\right]}}
\end{eqnarray}
and these are found above.

 The relation between the well dressed probability
amplitudes and the bare amplitudes is
\begin{equation}
D_{n,l,\pm}={1\over{\sqrt{2}}}(C_{e,n-1,l}\pm C_{g,n,l})
\end{equation}

Before turning to our results, let us recall the relations that
$g^{(2)}(\tau)$ must satisfy {\it if} the field can be described
by a classical stochastic process; if it has a positive definite
Glauber-Sudarshan P distribution,

\begin{eqnarray}
 g^{(2)}(0)&\geq& 1 \label{sp}\\
g^{(2)}(0+)&\geq& g^{(2)}(0)\label{pa}\\
|g^{(2)}(\tau)-1|&\leq& |g^{(2)}(\tau)-1|\label{ou}
\end{eqnarray}

Violations of all three of these inequalities has been observed in
CQED systems \cite{expt1,expt2}

\section{Results for intensity correlations}

As the system has a steady-state wave function in the steady state, we may write
\begin{equation}
 g^{(2)}(\tau)={{\langle\Psi_C(\tau)|a^{\dagger}a|\Psi_C(\tau)\rangle}\over{\langle a^{\dagger}a\rangle_{SS}}}.
 \end{equation}
 where we define
 \begin{equation}
 |\psi_C\rangle={{a|\psi_{SS}\rangle}\over{|a|\psi_{SS}\rangle|}}={{a|\psi_{SS}\rangle}\over{\sqrt{\langle|\psi_{SS}|a^{\dagger}a|\psi_{SS}\rangle|}}}
 \end{equation}
   In Fig. 3 we plot $g^{(2)}(0)-1$ for an initial state $\mid
\Psi_g\rangle$ where there is equal population in the $\mid 0,
l,g\rangle$ states for $l=0,l_{max}$. As more states are involved,
we find that the antibunching goes away. This is due to the fact
that the two single-photon vibronic ladders have a different
frequency spacing than the ground state vibronic levels, and is
consistent with the effect of detunings on the photon
statistics\cite{thy}. Involving more $l$ states makes the width of
the state larger, increasing $\Delta x$ for the center of mass
wave function of the atom. The antibunching also goes away if we
just prepare the system in a particular higher $l$ state $\mid
0,l_0,g\rangle$. The optimum state would seem to be the ground
state of the bare vibronic potential.

\begin{figure}
   \begin{center}
   \begin{tabular}{c}
   \includegraphics[height=9cm]{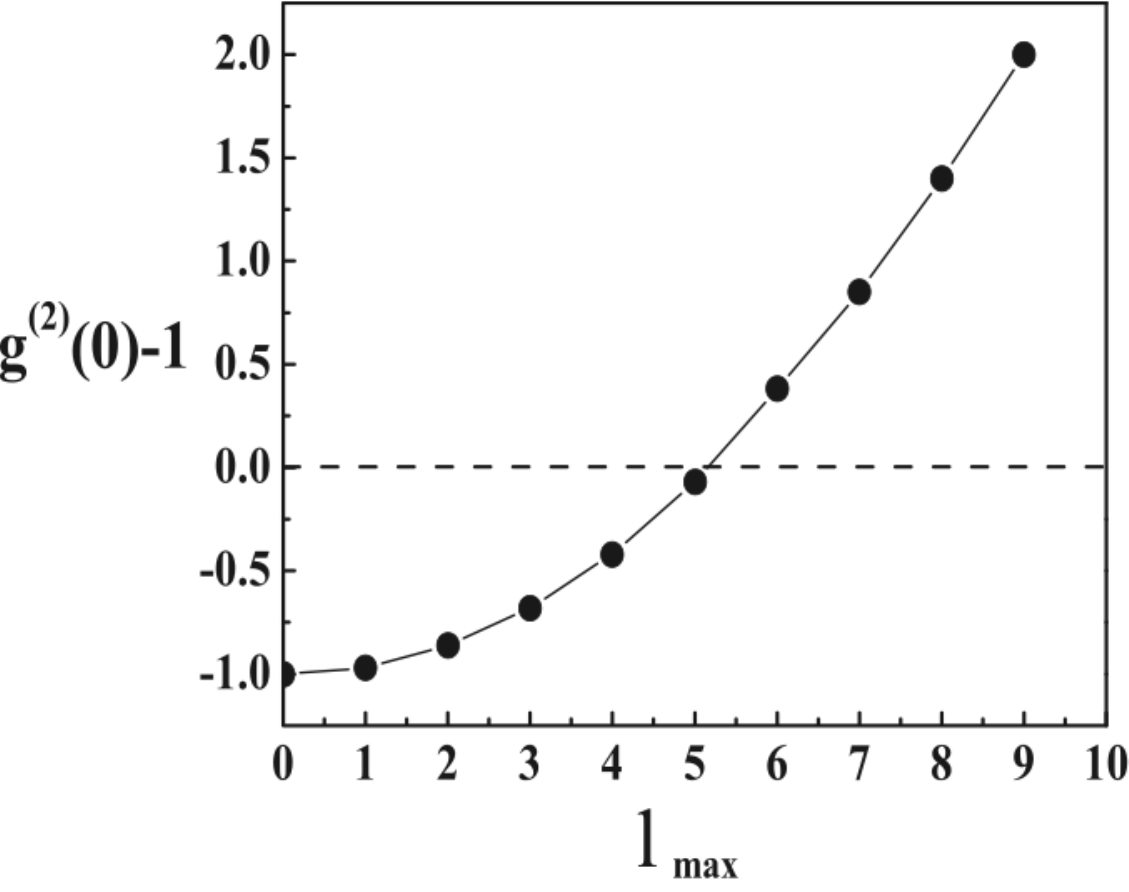}
   \end{tabular}
   \end{center}
   \caption[example]
%>>>> use \label inside caption to get Fig. number with \ref{}
%   { \label{fig:example}
{$g^{(2)}(0)-1$ vs. $N_{max}$, the highest occupied phonon number
}
   \end{figure}
Instead of just assigning values to the probability amplitudes (subject to normalization) we can specify the center of mass wave function and calculate the probability amplitudes via $D_{g,0,1}(0)=\langle \Psi_{CM}|l \rangle$. If we choose the center of mass wave function to be a Gaussian of width $\sigma$, we can calculate these amplitudes easily using
\begin{eqnarray}
D_{g,0,l}&=&\langle \Psi_{CM}|l \rangle\nonumber \\
&=&A_{n,l}\int^{\infty}_{\infty}e^{-y^2/2\sigma^2}e^{-y^2/2\sigma_0^2}H_l(y)dy
\end{eqnarray}
where $y=x/\sigma_0$ and $\sigma_0=\sqrt{\hbar/m\Omega_{0,l}}$ is the width of the ground state of the vibronic potential, and the normalization constant is $A_{n,l}=(m\Omega_{0,l}/\pi\hbar)^{1/4}/\sqrt{2^n n!}$.
\begin{figure}
   \begin{center}
   \begin{tabular}{c}
   \includegraphics[height=9cm]{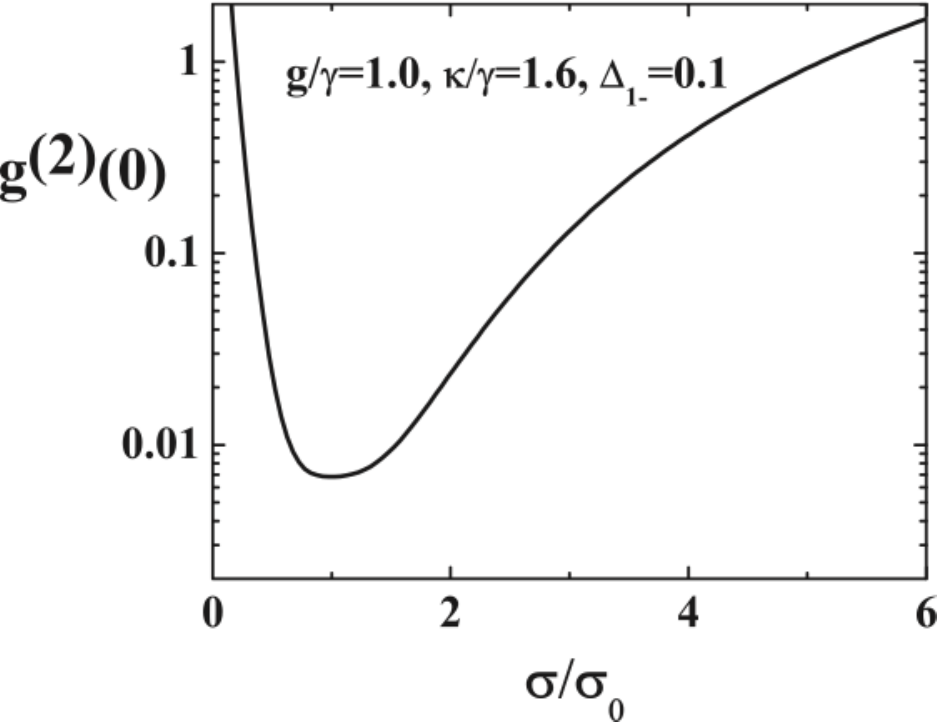}
   \end{tabular}
   \end{center}
   \caption[example]
%>>>> use \label inside caption to get Fig. number with \ref{}
%   { \label{fig:example}
{$g^{(2)}(0)-1$ vs. $\sigma/\sigma_0$, the relative width of a Gaussian center of mass wave function
}
   \end{figure}
In Fig. 4 we show a plot of $g^{(2)}(0)$ as a function of $\sigma/\sigma_0$ for parameters for which there is nearly perfect antibunching in the absence of an external potential. We see that there is a relatively wide region where the antibunching persists, but for $\sigma/\sigma_0$ less than $0.2$ or larger than $4$, the antibunching vanishes completely. This can be understood by considering that a Gaussian wave function is superposition of various vibronic states, and that population of higher excited vibronic states is deleterious to antibnching. Only when $\sigma/\sigma_0\approx 1$ do we have a center of mass wave function that has population predominantly in the ground state
In Fig. 5, we show a plot of $g^{(2)}(\tau)$  for $ g/\gamma =
2$, $\kappa/\gamma = 5$, $\Delta_{1,+}/\gamma = 0.1$. Fig. 5a is
for the atom initially in the ground state of the potential. We
see that $g^{(2)}(0)$ is about $1.1$. Classically, $g^{(2)}(\tau)$
could not then go below $0.9$, but here it goes to zero. We refer
to this as an undershoot. In Fig. 5b, we exhibit $g^{(2)}(\tau)$
for an equal admixture of the ground state and fifth excited
state. Here $g^{(2)}(0)$ is $4$, and hence the fact that
$g^{(2)}(\tau)$ is later zero is not nonclassical. The physical
reason for this can be traced back to the fact that an atom in an
excited state of the external potential is essentially detuned
from resonance. With $\Delta_{1,+}/\gamma=0.1$, the detuning
$\Delta_{5,+}/\gamma=0.5$. Previous work has shown that usually a
detuning of half a linewidth is quite deleterious to nonclassical
effects in $g^{(2)}(\tau)$. In Fig. 5c, we have results for what
we refer to as a pseudo-Boltzmann. Here we populate 20 vibronic
levels of the external potential at a "temperature" of 3mK. There
is no decoherence associated with this distribution, i.e. all the
off-diagonal matrix elements are not zero. This essentially
results in a distribution over populations with small population in the first excited state,even less in the
second, and so on. Here we see that with most of the population in
the ground state, we essentially have the ground state result. In
Fig. 5d, we show $g^{(2)}(\tau)$ for an equal population in all
$20$ vibronic states. Here we see large photon bunching, and no
nonclassical effects at all. This can be understood in terms of
detunings of the various atomic states; this type of distribution
over vibronic states would correspond to an atom more localized
than the ground state of the external potential. Hence we see that
localizing the atom too much results in a large spread in momentum
states that destroys the nonclassical effects.
\begin{figure}
 \centering
 \includegraphics[height=9cm]{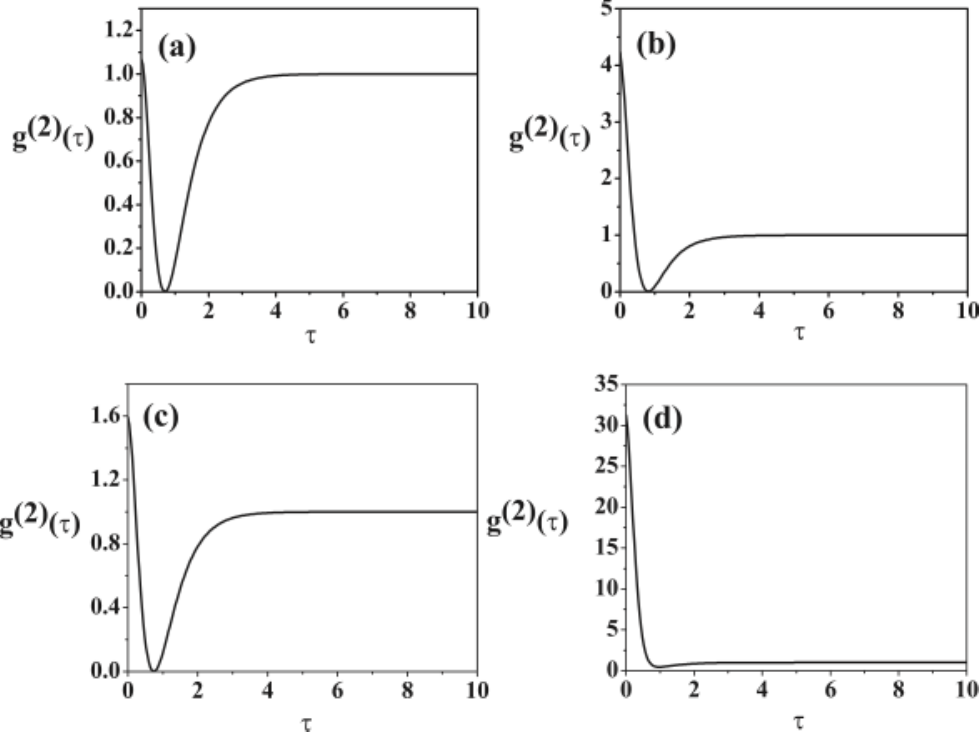}
 \caption{Plots are $g_{TT}^{(2)}(\tau)$ for $ g/\gamma = 2$, $\kappa/\gamma = 5$, $\Delta_{1,+}/\gamma = 0.1$, and (a) $|0\rangle$ only,  (b) $\frac{1}{\sqrt{2}}\left[ |0
 \rangle + |5 \rangle \right]$,  (c) Pseudo-Boltzmann, and (d) for 20 states with equal population. } \label{Fig5}
\end{figure}
In Fig. 6, we see that the fluorescent intensity correlations are
relatively insensitive to the choice of atomic center of mass wave
function, in that $g^{(2)}(0)$ is $0$ due to the nature of
single-atom fluorescence.
\begin{figure}
 \centering
\includegraphics[height=9cm]{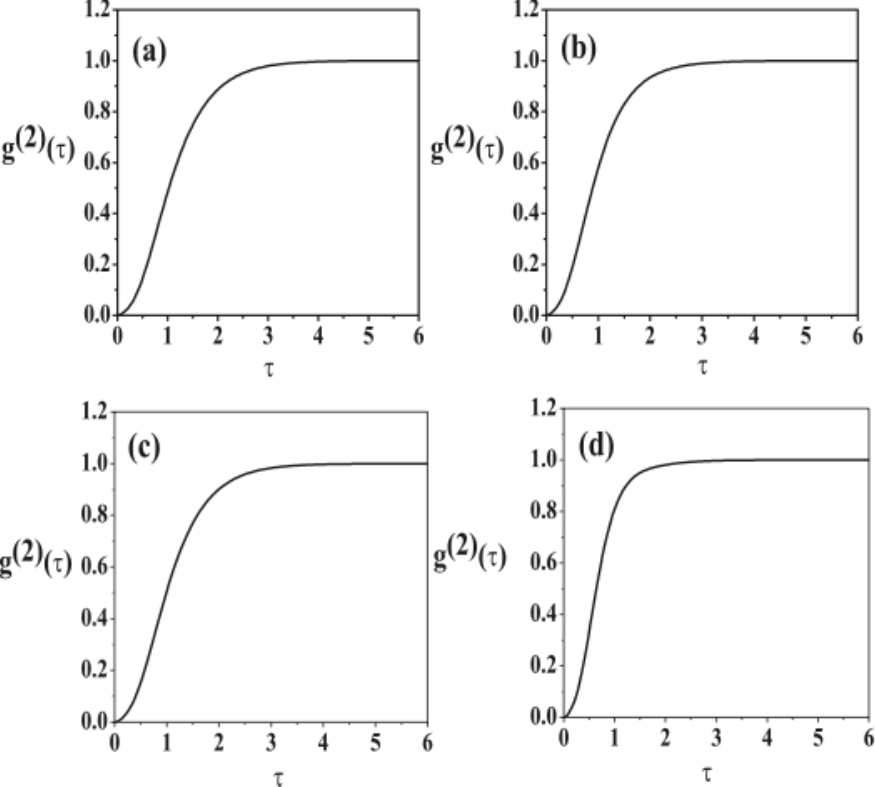}
 \caption{Plots are $g_{FF}^{(2)}(\tau)$ for $ g/\gamma = 2$, $\kappa/\gamma = 5$, $\Delta_{1,+}/\gamma = 0.1$, and (a) $|0\rangle$ only,  (b) $\frac{1}{\sqrt{2}}\left[ |0
 \rangle + |5 \rangle \right]$,  (c) Pseudo-Boltzmann, and (d) for 20 states with equal population. } \label{Fig6}
\end{figure}
In Fig. 7 we examine $g^{(2)}(\tau)$ for parameters where the
transmitted intensity correlation function $g^{(2)}(0)=0.0$. We
see that for a superposition of ground and fifth excited states,
we still have nonclassical effects, as $g^{(2)}(0) \leq 1$. The
initial slope of $g^{(2)}(\tau)$ is negative though, which is not
nonclassical. For the pseudo-Boltzmann distribution, we see both
types of nonclassical behaviors. In the case of equal population
over 20 vibronic states, there is no nonclassical behavior at all.
\begin{figure}
 \centering
\includegraphics[height=9cm]{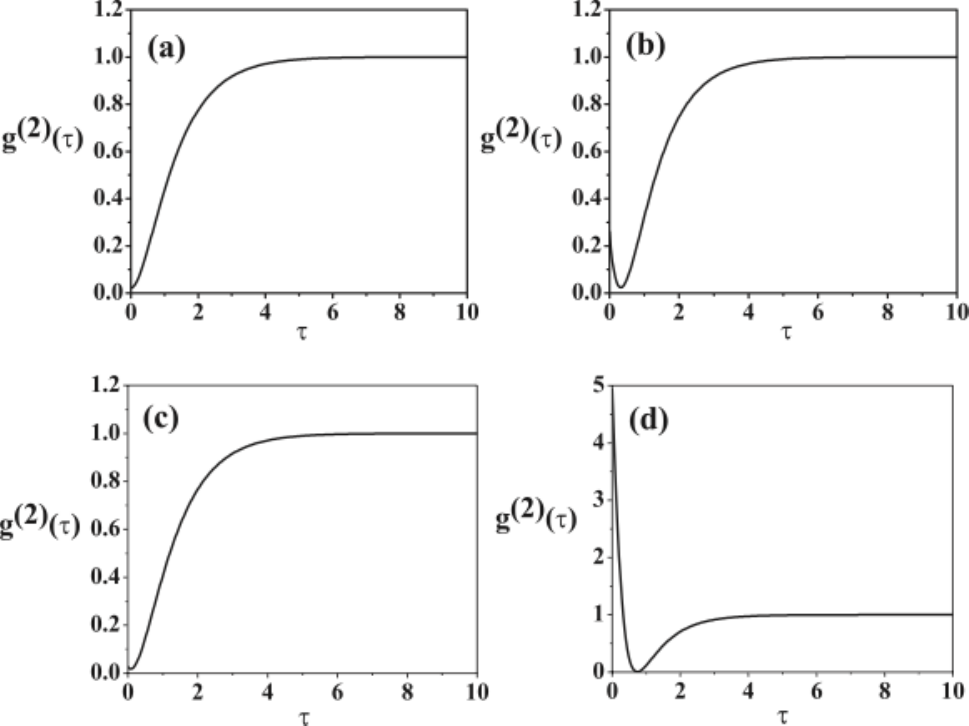}
 \caption{Plots are $g_{TT}^{(2)}(\tau)$ for $ g/\gamma = 2.2$, $\kappa/\gamma = 10$, $\Delta_{1,+}/\gamma = 0.1$, and (a) $|0\rangle$ only,  (b) $\frac{1}{\sqrt{2}}\left[ |0
 \rangle + |5 \rangle \right]$,  (c) Pseudo-Boltzmann, and (d) for 20 states with equal population. } \label{Fig7}
\end{figure}
In Fig. 8, we look at a case where there is strong coupling, but
no nonclassical behavior in the ground state case. We do have
strong vacuum-Rabi oscillations. For an admixture of states, we
see a beat frequency in the oscillations. The pseudo-Boltzmann
case again is very similar to the ground state case. The
oscillations are almost completely washed out when we have equal
population in 20 vibronic states.
\begin{figure}
 \centering
\includegraphics[height=9cm]{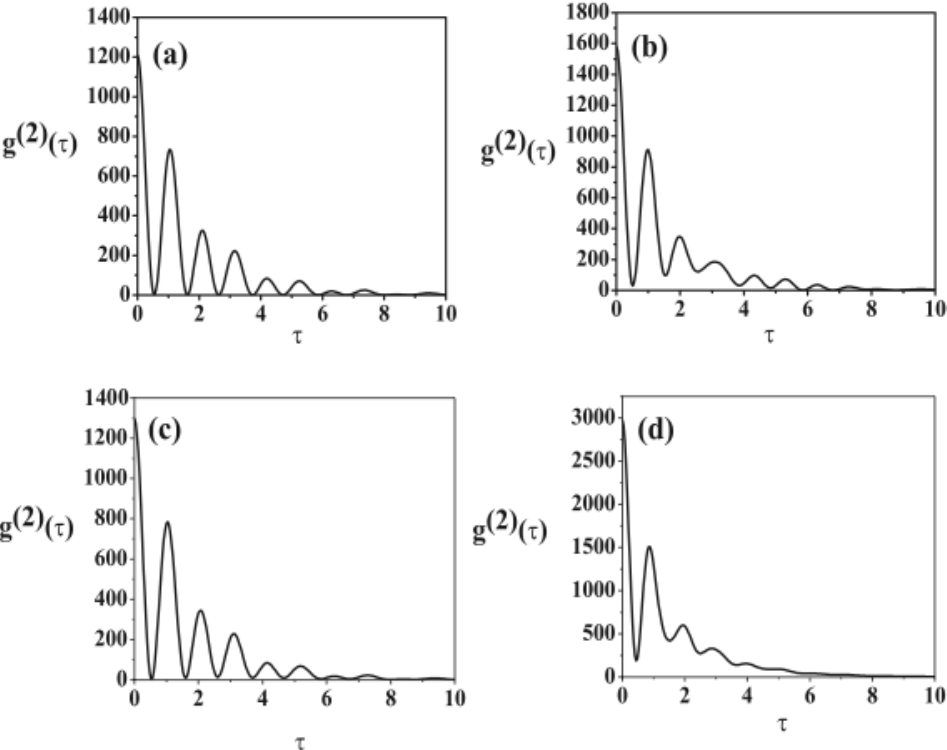}
\caption{Plots are $g_{TT}^{(2)}(\tau)$ for $ g/\gamma = 3$,
$\kappa/\gamma = .1$, $\Delta_{1,+}/\gamma = 0.1$, and (a)
$|0\rangle$ only,  (b) $\frac{1}{\sqrt{2}}\left[ |0
 \rangle + |5 \rangle \right]$,  (c) Pseudo-Boltzmann, and (d) for 20 states with equal population. } \label{Fig8}
\end{figure}
 In
Fig. 9 we again look at a situation where the ground state case
shows strong vacuum-Rabi oscillations as well as all three
nonclassical behaviors; $g^{(2)}(0) \leq 1$, the initial slope is
positive, and there is an overshoot violation. The latter refers
to $g^{(2)}(\tau)$ violating the upper limit of the inequality in
\ref{ou}.
\begin{figure}
 \centering
 \includegraphics[height=9cm]{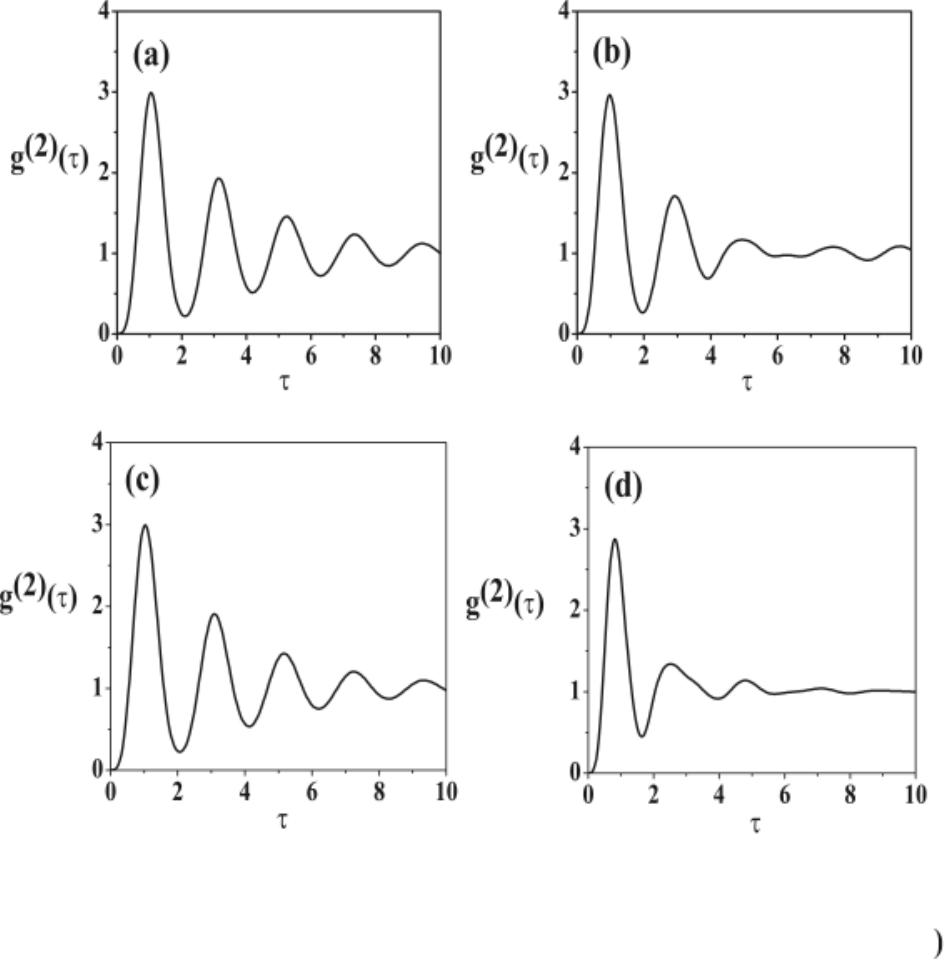}
 \caption{Plots are $g_{FF}^{(2)}(\tau)$ for $ g/\gamma = 3$, $\kappa/\gamma = .1$, $\Delta_{1,+}/\gamma = 0.1$, and (a) $|0\rangle$ only,  (b) $\frac{1}{\sqrt{2}}\left[ |0
 \rangle + |5 \rangle \right]$,  (c) Pseudo-Boltzmann, and (d) for 20 states with equal population.  } \label{Fig9}
\end{figure}
In Fig. 10 we have a situation where we only have an overshoot
violation in the ground state case. This violation vanishes in the
case of a superposition of ground and fifth excited state, as well
as for an equal population of 20 vibronic states. In Figure 11, we
examine the effects of increasing spacing between the vibronic
levels. To this point we have dealt with detunings on the order of
$0.1$ linewidths.
\begin{figure}
 \centering
 \includegraphics[height=9cm]{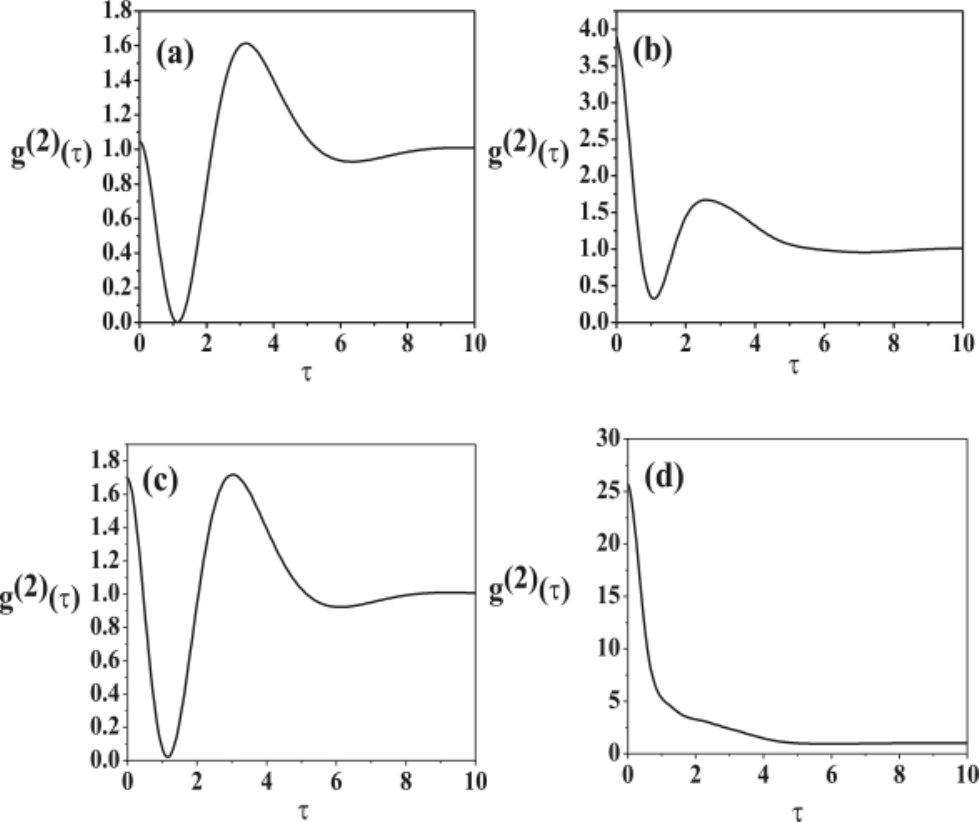}
 \caption{Plots are $g_{TT}^{(2)}(\tau)$ for $ g/\gamma = 1$, $\kappa/\gamma = .77$, $\Delta_{1,+}/\gamma = 0.1$, and (a) $|0\rangle$ only,  (b) $\frac{1}{\sqrt{2}}\left[ |0
 \rangle + |5 \rangle \right]$,  (c) Pseudo-Boltzmann, and (d) for 20 states with equal population.  } \label{Fig10}
\end{figure}
In Fig. 12 we can see that increasing the detunings allows us to
see a larger effect due to the beat frequency. Changing the
detuning to $0.3$ and $0.5$ of $\gamma$, we see that the initial
slope is not nonclassical, but we still have $g^{(2)}(0) \leq 1$,
and there is an undershoot violation. So the nature of the
nonclassicality is not changed. At a detuning of $2.0$, we still
have an undershoot violation as well as evidence of oscillations
at the beat frequency.
\begin{figure}[here]
 \centering
\includegraphics[height=9cm]{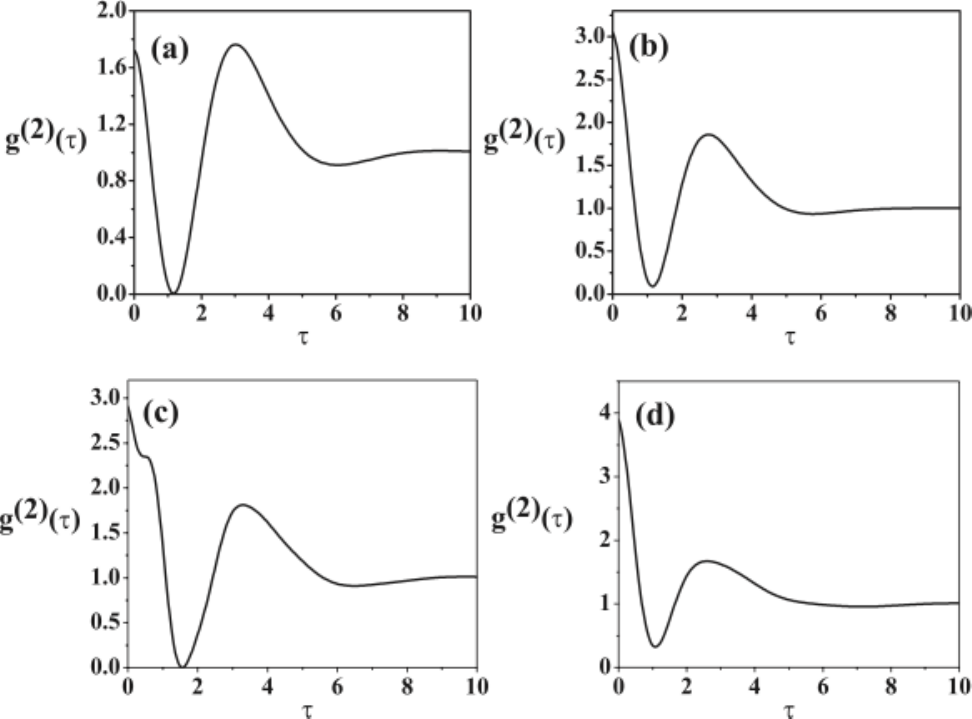}
  \caption{Plots are $g_{TT}^{(2)}(\tau)$ for $ g/\gamma = 1$, $\kappa/\gamma = 1.6$.  All trials use $\frac{1}{\sqrt{2}}\left[ |0
 \rangle + |5 \rangle \right]$ as the vibrational state distribution with  (a) $\Delta_{1,+}/\gamma =
 0.1$,
 (b) $\Delta_{1,+}/\gamma = 0.3$,  (c) $\Delta_{1,+}/\gamma = 2.0$,  (d) $\Delta_{1,+}/\gamma = 0.5$. } \label{Fig11}
\end{figure}
In Figure 12, we examine the effects of increasing spacing between
the vibronic levels. In this case we have antibunching, a
violation of inequality in Eq.(\ref{sp}). Changing the
$\Delta_{1,+}/\gamma$ to $0.3$ and $0.5$, we see that the initial
slope is not nonclassical, but we still have $g^{(2)}(0) \leq 1$,
and there is an undershoot violation\cite{expt2}. So the nature of
that nonclassicality is not changed. At a detuning of $2.0$,(Fig.
5c) we still have an undershoot violation as well as evidence of
oscillations at the beat frequency.

\begin{figure}[here]
 \centering
\includegraphics[height=9cm]{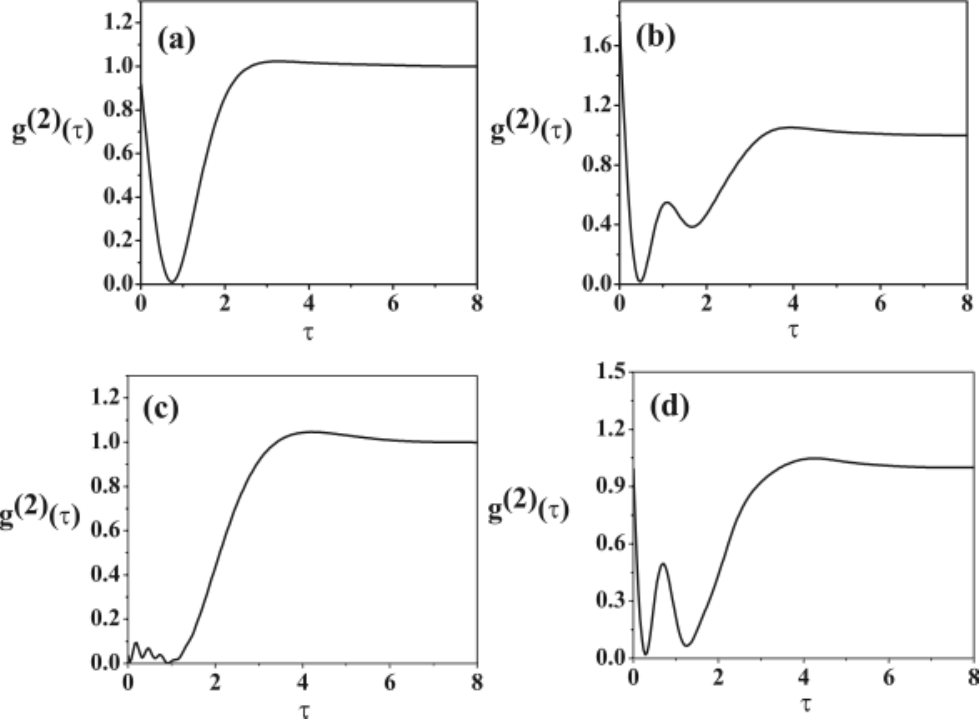}
  \caption{Plots are $g_{TT}^{(2)}(\tau)$ for $ g/\gamma = 1$, $\kappa/\gamma = 1.6$.  All trials use $\frac{1}{\sqrt{2}}\left[ |0
 \rangle + |5 \rangle \right]$ as the vibrational state distribution with  (a) $\Delta_{1,+}/\gamma =
 0.1$,
 (b) $\Delta_{1,+}/\gamma = 0.3$,  (c) $\Delta_{1,+}/\gamma = 2.0$,  (d) $\Delta_{1,+}/\gamma = 0.5$. } \label{Fig12}
\end{figure}

\section{Wave-Particle Correlations}
Recently, Carmichael his co-workers have introduced a
new intensity-field correlation function $h_{\theta}(\tau)$ that
is of great interest \cite{HT1,HE1}.  Because
$h_{\theta}(\tau)$ is an intensity-field correlation function, it
takes the general form
\begin{equation}
h_{\theta}(\tau)={{\langle I(0) E(\tau)\rangle}\over{\langle
I\rangle\langle E\rangle}},
\end{equation}
and for a quantized field, this becomes
\begin{equation}\label{htheta}
h_{\theta}(\tau)={{\langle {a}^{\dagger}(0)
a_{\theta}(\tau)a(0)\rangle}\over{\langle
{a}^{\dagger}a\rangle\langle a_{0} \rangle}},
\end{equation}
where we have, like for $g^{(2)}(\tau)$, exploited normal and time
ordering, and we have used the quantum mechanical field quadrature
operator :
\begin{equation}\label{quadop}
{a}_{\theta}={1\over2}\left(ae^{-\imath \theta}
+{a}^{\dagger}e^{\imath \theta}\right).
\end{equation}
In Eq. (\ref{quadop}), $\theta$ is the phase of the local
oscillator (LO) with respect to the average signal field. We see
that with the $a$ acting to the right, and the ${a}^{\dagger}$
acting to the left at $t=0$, a collapsed state is prepared, the
collapse being a photon loss from the field, corresponding to a
detection event. Then at $t=\tau$ one measures $\langle
a_{\theta}\rangle$ {\it conditioned} on the previous detection.
This differs from a direct measurement of $\langle
a_{\theta}\rangle$ with no conditioning. An ensemble average of
the latter measurements (necessary to get a good signal to noise
ratio) would yield zero due to phase fluctuations. The conditioned
BHD measurement essentially looks at members of the ensemble with
the same phase, a phase that is set by the photodetection.

As with other correlation functions, like the second-order
intensity correlation function $g^{(2)}(\tau )$, restrictions can
be placed on $h_{\theta}(\tau)$ {\it if} there is an underlying
 positive definite probability distribution function for amplitude
 and phase of the electric field, i.e. that the field is classical
 albeit stochastic. By ignoring third-order moments (a Gaussian
 fluctuation assumption that is valid for weak fields), one finds
 \begin{equation}\label{howfluc}
 h_{\theta}(\tau)=1
 +2{{\langle:\Delta{a}_{\theta}(0)\Delta{a}_{\theta}(\tau):\rangle}\over
 {\langle \Delta{a}^{\dagger}\Delta{a}\rangle}},
 \end{equation}
 and we see that the intensity-field correlation function is
 connected to the spectrum of squeezing \cite{HT1}
 \begin{equation}
 S_{\theta}(\omega)\propto \int_0^{\infty}d\tau \cos (\omega \tau
 )\left[h_{\theta}(\tau)-1\right].
 \end{equation}
 From this, it has been shown that the Schwartz inequality would
 yield
 \begin{equation}
 0\leq h_{\theta}(0)-1 \leq 1,\\
\end{equation}
and more generally
\begin{equation}\label{SIH}
\mid h_{\theta}(\tau)-1\mid \leq \mid h_{\theta}(0)-1 \mid \leq 1.
\end{equation}
Whenever there is squeezing, these inequalities do not hold for $
h_{\theta}(\tau)$. Giant violations of these inequalities have
been predicted for an optical parametric oscillator, and a group
of $N$ atoms in a driven optical cavity, and have been recently
observed in the cavity QED system \cite{HE1}.

Now consider the following quantity
\begin{equation}
\langle IE\rangle ^2\leq \langle I\rangle^2\langle E\rangle^2
\end{equation}
After some algebra we find
\begin{eqnarray}
h^2_{0}(0)&=&{{\langle IE\rangle ^2}\over{\langle I\rangle^2\langle E\rangle^2}}\nonumber \\
&\leq&{{\langle I^2\rangle\langle E^2\rangle}\over{\langle E\rangle^2\langle I\rangle^2}}\nonumber \\
&\leq&{{\langle I\rangle}\over{\langle E\rangle^2}}g^{(2)}(0)\\
&\leq&g^{(1)}(0)g^{(2)}(0)\label{hg2ineq}
\end{eqnarray}
In the absence of an external potential $g^{(2)}(0)=|C_{g,1}^{CT}/C_{g,1}^{ss}|^2=h^2_{0}(0)$. As $h_{0}(0)$ will be nonclassical above $2$, we must have bunching to see nonclassical behavior in the conditioned fields. Also in the  system considered here , we would have $h^2\leq g^{(2)}$; when we include an optical lattice
we have
\begin{eqnarray}
h_{0}(0)&=&{{\sum_k C^c_{1,g,k}}\over{\sum_k C^{ss}_{1,g,k}}}\\
g^{(2)}(0)&=&{{\sum_k |C^c_{1,g,k}|^2}\over{\sum_k |C^{ss}_{1,g,k}|^2}}\\
|h_{0}(0)|^2&=&{{|\sum_k C^c_{1,g,k}|^2}\over{|\sum_k C^{ss}_{1,g,k}|^2}}
\end{eqnarray}
Just looking at the numerator, for two vibronic modes {k values}, we would violate Eq. (\ref{hg2ineq}).

As with $g^{(2)}(\tau)$, we obtain an analytic solution using the quantum trajectory
method, and again we look at weak driving fields. We find
\begin{equation}
\langle {a}^{\dagger}(0) a_{\theta}(\tau)a(0)\rangle=\langle
\psi_c \mid a_{\theta}\mid \psi_c \rangle,
\end{equation}
where $\mid \psi_c\rangle$ is the collapsed state produced by the
photodetection event, as in the case of $g^{(2)}(\tau)$. Once
again we need only keep the states with two or less excitations
(total in the cavity mode or internal energy) for weak driving
fields. The result is that
\begin{eqnarray}
h_{\theta}(\tau)&=&{{\langle n\rangle_{SS}\langle {a}_\theta
(\tau) \rangle_{CT}}\over{{\langle n\rangle_{SS}\langle {a}_\theta
(\tau) \rangle_{SS}}}}\nonumber\\ &=&{{\langle {a}_\theta (\tau)
\rangle_{CT}}\over{\langle {a_0 (\tau) \rangle_{SS}}}}.
\end{eqnarray}
The expectation value of the field quadrature operator is given by
\begin{equation}
\langle{a}_{\theta}\rangle
=\sum_{n,l}\left(\sqrt{n}C^*_{n,l}C_{n-1,l}e^{-\imath\theta}
+\sqrt{n+1}C^*_{n,l}C_{n+1,l}e^{\imath\theta} \right).
\end{equation}
In the weak field limit we have
\begin{equation}
\langle{\hat{a}}_{\theta}\rangle
=\sum_{l}\left(C^*_{1,l}C_{0,l}e^{-\imath\theta}+
C^*_{0,l}C_{1,l}e^{\imath\theta} \right).
\end{equation}

So finally then, for weak fields we have
\begin{equation}
h_{\theta}(\tau)=\frac{\sum_{l}\left(C^{CT}_{1,l}C^{CT}_{0,l}e^{-\imath\theta}+
{C^{CT}}^*_{0,l}C^{CT}_{1,l}e^{\imath\theta} \right)}
{\sum_{l}\left(C^{SS*}_{1,l}C^{SS}_{0,l}+
C^{SS*}_{0,l}C^{SS}_{1,l} \right)}.
\end{equation}

For the fluorescent field, we have
\begin{equation}
h_{\theta}^{FF}(\tau)={{\langle \sigma_{\theta} (\tau)
\rangle_{CF}}\over{\langle \sigma_0 (\tau) \rangle_{SS}}}.
\end{equation}
which in terms of probability amplitudes is
\begin{equation}
h_{\theta}^{FF}(\tau)={{2Re{\sum_{l}C^{CF}_{0,e,l}(\tau)C_{0,g,l}^{CF}(\tau)}e^{i\theta}
}\over {{\sum_{l}C^{SS}_{0,e,l}C_{0,g,l}^{SS}}}}.
\end{equation}
In Fig. 14 we plot $h_{\theta}^{TT}$ for $ g/\gamma = 2$,
$\kappa/\gamma = 5$, for the same choice of four states we have
used. We see that in the case of a highly localized atom (equal
probability of 20 vibronic levels) the nonclassical nature of
$h_{\theta}^{TT}$ is actually enhanced. In the case of an
admixture of ground and fifth excited states, the behavior of
$h_{\theta}^{TT}$ is relatively unchanged from the ground state
case. This is due to the insensitivity of $h_{\theta}^{TT}$ to
detunings in the weak coupling limit. In the strong coupling
regime, as shown in Fig. 15, we see the same general behavior,
although for the case of 20 equal populations we do see some
dephasing of the vacuum-Rabi oscillations, due to the detunings of
the various levels involved. Similar behavior is seen in the case
of $h_{\theta}^{FF}$ as shown in Figs. 16 and 17. Note that
$h_{\theta}^{FF} (0)=0.0$, reflecting the fact that after
spontaneous emission, the dipole field envelope vanishes. In Fig.
18 we change the level spacing. We see that for increasing
vibronic level spacing the nature of the
nonclassicality persists, but there is evidence of the beat
frequency between subsequent vibronic levels.
\begin{figure}
 \centering
 \includegraphics[height=9cm]{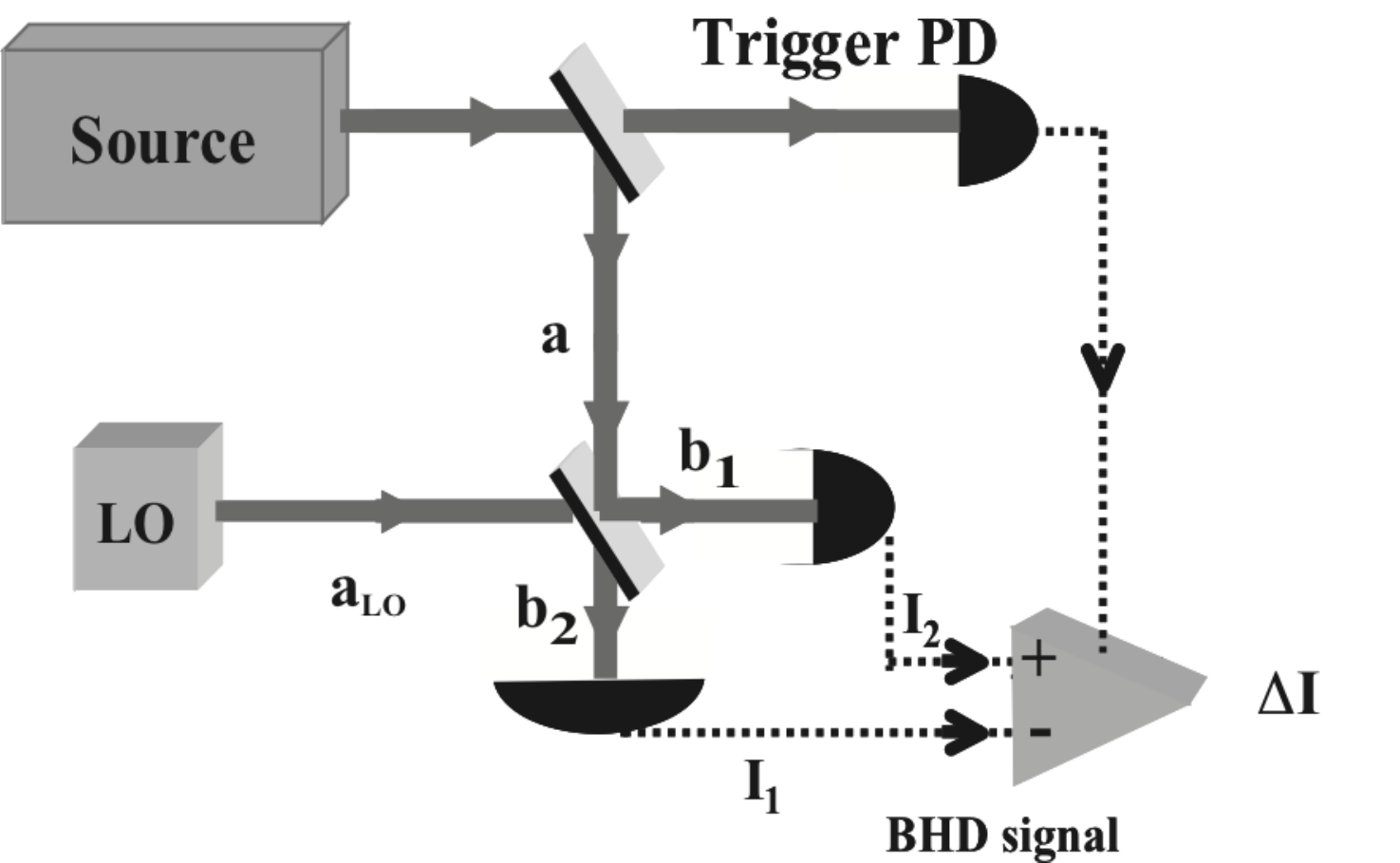}
 \caption{This is a common experimental setup for measuring $h_{\theta}(\tau)$.  In this figure,
  the source would be either the transmitted or fluoresced portion of the field.  LO denotes Local
  Oscillator, a } \label{fig:Htheta}
\end{figure}

\begin{figure}
 \centering
\includegraphics[height=9cm]{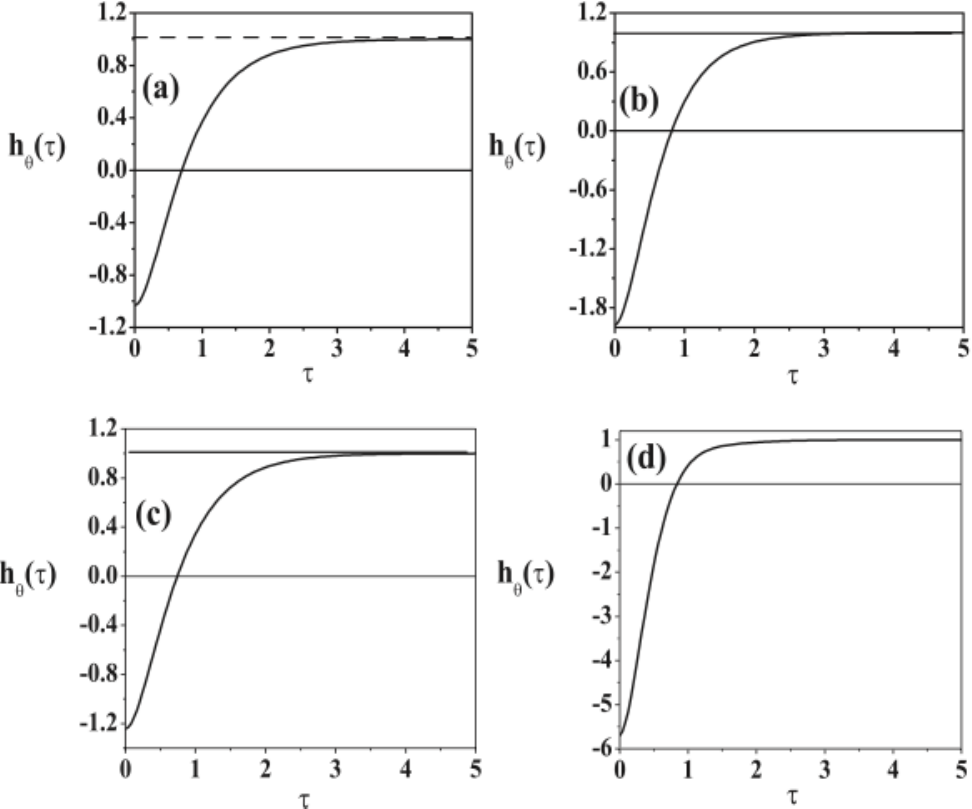}
\caption{Plots are $h^{TT}_{\theta}(\tau)$ for $ g/\gamma = 2$,
$\kappa/\gamma = 5$
 , $\Delta_{1,+}/\gamma = 0.1$.  (a) $|0\rangle$ only.  (b) $\frac{1}{\sqrt{2}}\left[ |0
 \rangle + |5 \rangle \right]$.  (c) Pseudo-Boltzmann. (d) All states equal population. }\label{Fig14}
\end{figure}

\begin{figure}
 \centering
\includegraphics[height=9cm]{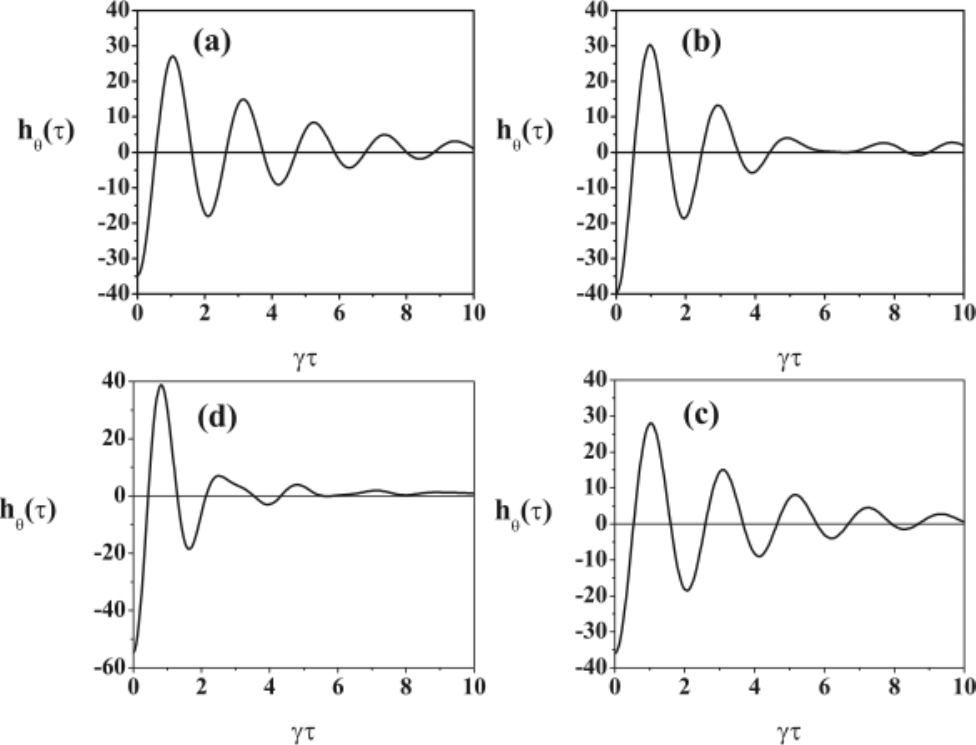}
\caption{Plots are $h^{TT}_{\theta}(\tau)$ for $ g/\gamma = 3$,
$\kappa/\gamma = .1$
 , $\Delta_{1,+}/\gamma = 0.1$.  (a) $|0\rangle$ only.  (b) $\frac{1}{\sqrt{2}}\left[ |0
 \rangle + |5 \rangle \right]$.  (c) Pseudo-Boltzmann. (d) All states equal population. }\label{fig15}
\end{figure}

\begin{figure}
 \centering
\includegraphics[height=9cm]{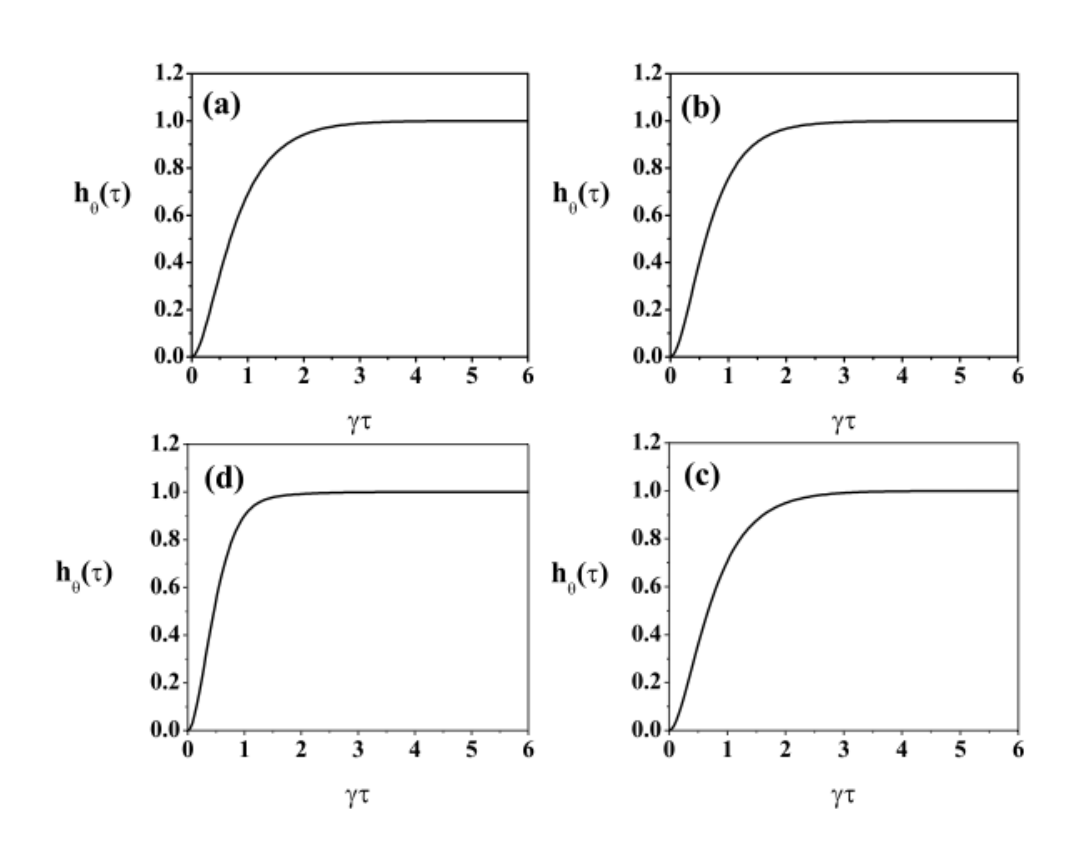}
\caption{Plots are $h^{TT}_{\theta}(\tau)$ for $ g/\gamma = 3$,
$\kappa/\gamma = .1$
 , $\Delta_{1,+}/\gamma = 0.1$.  (a) $|0\rangle$ only.  (b) $\frac{1}{\sqrt{2}}\left[ |0
 \rangle + |5 \rangle \right]$.  (c) Pseudo-Boltzmann. (d) All states equal population. }\label{fig16}
\end{figure}

\begin{figure}
\includegraphics[height=9cm]{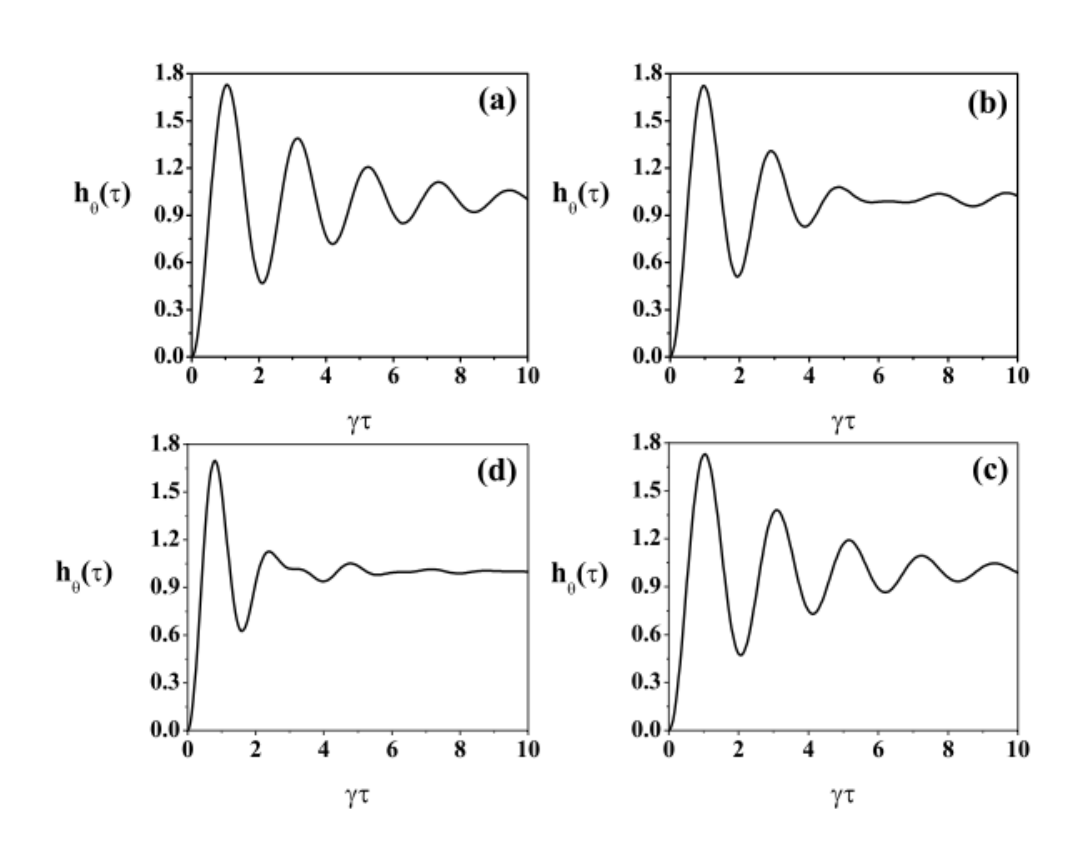}
 \caption{Plots are $h^{FF}_{\theta}(\tau)$ for $ g/\gamma = 1$, $\kappa/\gamma = .77$
 , $\Delta_{1,+}/\gamma = 0.1$.  (a) $|0\rangle$ only.  (b) $\frac{1}{\sqrt{2}}\left[ |0
 \rangle + |5 \rangle \right]$.  (c) Pseudo-Boltzmann. (d) All states equal population. }\label{Fig17}
\end{figure}

\begin{figure}
\includegraphics[height=9cm]{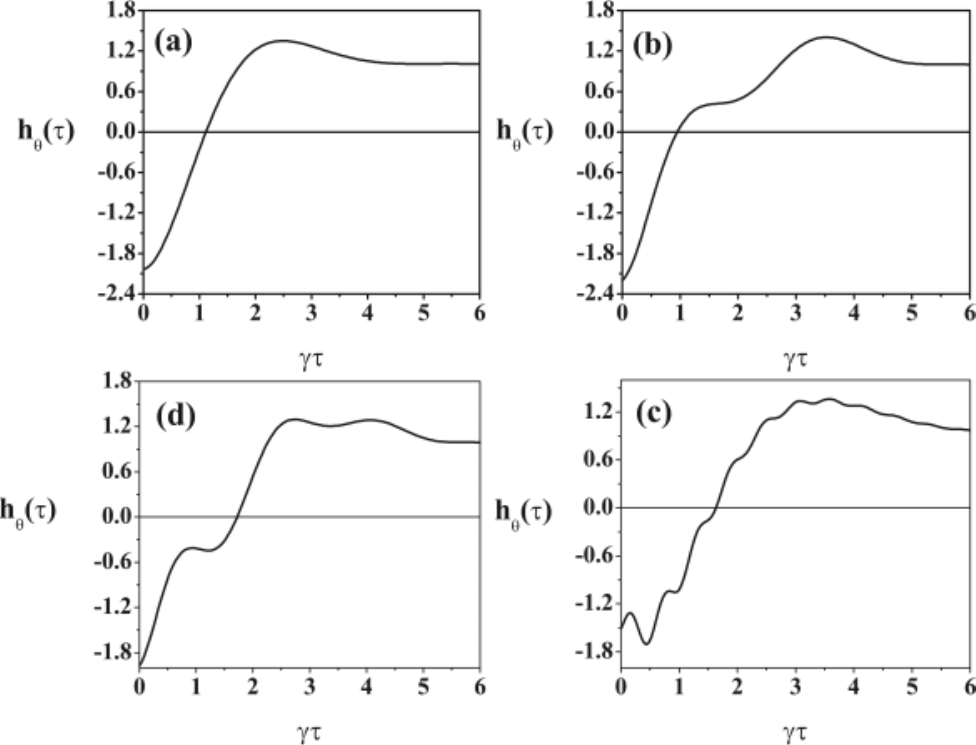}
 \caption{Plots are $h^{TT}_{\theta}(\tau)$ for $ g/\gamma = 1$, $\kappa/\gamma = .77$
 , $\Delta_{1,+}/\gamma = 0.1$.  (a) $|0\rangle$ only.  (b) $\frac{1}{\sqrt{2}}\left[ |0
 \rangle + |5 \rangle \right]$.  (c) Pseudo-Boltzmann. (d) All states equal population. }\label{Fig18}
\end{figure}

\section{Conclusion} We have considered the photon statistics of a
cavity QED system while including quantized center of mass motion
along the cavity axis. In the limit of weak driving fields we have
found analytic results for intensity correlations of the
transmitted and fluorescent fields; as well as for the
cross-correlations between the transmitted and fluorescent
intensities. We find that for intensity correlations for the
transmitted field, having a significant population outside the
ground vibronic level is deleterious to sub-Poissonian statistics,
photon antibunching, and overshoot/undershoot violations. This is
explained due to the sensitivity of these nonclassical effects to
detunings between the atom-cavity system and the driving field. It
is found that significant population in vibronic levels that are
out of resonance by a half a linewidth is sufficient to severely
modify the results; a highly localized atom, spread over many
vibronic levels only exhibits nonclassical effects over a very
small parameter range. For the fluorescent intensity correlations,
we do not find such a sensitivity, this is due mainly to the
nature of single atom fluorescence where the atom can only emit
one photon at a time. The cross-correlations exhibit the assymetry
noted by Denisov et. al., and this asymmetry is not degraded
significantly by a distribution over vibronic levels.

We have also found analytic results for $h_{\theta}(\tau)$ for the
transmitted and fluorescent fields. There is no time asymmetry for
weak driving fields. The nonclassical behavior in
$h_{\theta}(\tau)$ is not generally degraded by a distribution
over vibronic levels; indeed it is sometimes enhanced.

Future work includes inclusion of 2- and 3-d external trapping
potentials, non-harmonic potentials, and pressing beyond the weak
field limit.

\end{document}